\documentclass[12pt]{article}
\usepackage{epsfig,amsmath,amssymb,mathrsfs}
\usepackage[utf8]{inputenc}
\usepackage[colorlinks=true,citecolor=blue,,linktocpage=true,linkcolor=blue,urlcolor=black]{hyperref}
\usepackage{mathtools,slashed}
\usepackage{array}
\usepackage{xcolor}
\usepackage[force]{feynmp-auto}
\usepackage{caption}
\usepackage{subcaption}

  {\end{list}}%

\makeatletter
\renewcommand{\section}{\setcounter{equation}{0}\@startsection
 {section}%
 {1}%
 {0pt}%
 {-1\baselineskip}%
 {0.4\baselineskip}%
 {\bfseries\large}}%
\renewcommand{\subsection}{\@startsection
 {subsection}%
 {2}%
 {0pt}%
 {-0.75\baselineskip}%
 {0.2\baselineskip}%
 {\bfseries}}%
\renewcommand{\subsubsection}{\@startsection
 {subsubsection}%
 {3}%
 {0pt}%
 {-0.5\baselineskip}%
 {0.1\baselineskip}%
 {\sc}}%
\makeatother

\DeclareMathAlphabet{\mathpzc}{OT1}{pzc}{m}{it}

\def\be{\begin{equation}}
\def\ee{\end{equation}}

\def\g5{\gamma_{5}}

\def\m{\mu}
\def\n{\nu}

\def\s{\sigma}

\def\id3k{\int\!\! \dfrac{d^3\!\vec{k}}{(2\pi)^3 2E(\vec{k})}}

\def\idkd{\int\!\! \dfrac{d^{d}k\!}{(2\pi)^{d}}}
\def\idqd{\int\!\! \dfrac{d^{d}q\!}{(2\pi)^{d}}}
\def\idpd{\int\!\! \dfrac{d^{d}p\!}{(2\pi)^{d}}}

\newcommand{\bea}{\begin{eqnarray}}
\newcommand{\eea}{\end{eqnarray}}
\newcommand{\beann}{\begin{eqnarray*}}
\newcommand{\eeann}{\end{eqnarray*}}
\newcommand{\ba}{\begin{array}}
\newcommand{\ea}{\end{array}}

 \def\g {\gamma}

\newcommand{\email}[1]{\href{mailto:#1}{\tt #1}}

\begin{document}

\rightline{\scriptsize{FT/UCM 121-2022}}
\vglue 50pt

\begin{center}

{\LARGE \bf Unimodular gravity and the gauge/gravity duality}\\
\vskip 1.0true cm
{\Large Jesus Anero$^{\dagger}$, Carmelo P. Martin$^{\dagger\dagger}$ }
\\
\vskip .7cm
{
	$\dagger$Departamento de F\'isica Te\'orica and Instituto de F\'{\i}sica Te\'orica (IFT-UAM/CSIC),\\
	Universidad Aut\'onoma de Madrid, Cantoblanco, 28049, Madrid, Spain\\
	\vskip .1cm
	{$\dagger\dagger$Universidad Complutense de Madrid (UCM), Departamento de Física Teórica and IPARCOS, Facultad de Ciencias Físicas, 28040 Madrid, Spain}
	
	\vskip .5cm
	\begin{minipage}[l]{.9\textwidth}
		\begin{center}
			\textit{E-mail:}
			\email{$\dagger$jesusanero@gmail.es},
			\email{$\dagger\dagger$carmelop@fis.ucm.es}.
			
		\end{center}
	\end{minipage}
}
\end{center}
\thispagestyle{empty}

\begin{abstract}
Unimodular gravity can be formulated so that transverse diffeomorphisms and Weyl transformations are symmetries of the theory. For this formulation of unimodular gravity,  we work out the two-point and three-point $h_{\mu\nu}$ contributions  to the on-shell classical gravity action in the leading approximation and for an Euclidean AdS background. We conclude that these contributions do not agree with those obtained by using General Relativity  due to IR divergent contact terms. The subtraction of  these IR divergent terms yields the same IR finite result for both unimodular gravity and General Relativity. Equivalence between unimodular gravity and  General Relativity with regard to the  gauge/gravity duality thus emerges in a non trivial way.
\end{abstract}

\vfill
\clearpage

\section{Introduction}

Unimodular gravity is a theory of gravity which puts the cosmological constant problem into a new perspective \cite{vanderBij:1981ym, Zee:1983jg, Buchmuller:1988wx, Henneaux:1989zc}, for the vacuum energy does not gravitate in that theory. In unimodular gravity the cosmological constant does not enter the classical action and thus it occurs as an integration constant in the classical theory \cite{vanderBij:1981ym, Zee:1983jg, Buchmuller:1988wx, Henneaux:1989zc}. At  the quantum level, the cosmological constant occurs as parameter of the background field when computing the on-shell perturbative background-field effective action \cite{Alvarez:2015sba} and as a property of boundary states when computing transition amplitudes between those states  \cite{Buchmuller:2022msj}.

In the current century, several issues have been studied over the years in connection with unimodular gravity --see \cite{Carballo-Rubio:2022ofy}, for a recent review. Let us mention just a few.  Unimodular gravity as one of the two sound theories with transverse-diffeomorphism invariance \cite{Alvarez:2006uu}.  How unimodular gravity arises from interacting gravitons  \cite{Barcelo:2014mua}. The quantization of unimodular gravity within the BRST formalism \cite{Alvarez:2015sba, Upadhyay:2015fna, Kugo:2022iob, Kugo:2022dui, Baulieu:2020obv, Buchmuller:2022msj}. Whether unimodular gravity and general relativity agree as effective quantum field theories \cite{Alvarez:2005iy, Fiol:2008vk, Bufalo:2015wda, Alvarez:2016uog, deLeonArdon:2017qzg, Gonzalez-Martin:2017fwz, Gonzalez-Martin:2018dmy, deBrito:2021pmw}. Asymptotic-safety analysis of unimodular gravity \cite{Eichhorn:2013xr, Saltas:2014cta, Eichhorn:2015bna, DeBrito:2019gdd, deBrito:2020rwu}. The formulation of unimodular supergravity \cite{Anero:2019ldx, Anero:2020tnl, Bansal:2020krz}. Sundry topics like the first order formalism \cite{Alvarez:2015oda} and the hamiltonian formalism \cite{Alvarez:2021fhp} as applied to unimodular gravity, and a massive version of the theory \cite{Alvarez:2018law}.

The gauge/gravity duality conjecture  states that a gravity theory in a $d+1$-dimensional space-time with boundary is equivalent to an appropriate gauge theory --with no gravity-- in its $d$ dimensional boundary. There is a wealth of evidence --see \cite{Ammon:2015wua, Nastase:2015wjb} and references therein-- that this conjecture holds for the pair of theories for which the duality was originally put forward \cite{Maldacena:1997re}, namely: type IIB superstring on $AdS_5\times S^5$ with $N$ units of flux on $S^5$, on the one hand, and ${\cal N}=4$ super-Yang-Mills for $SU(N)$ on four dimensional Minkowski space-time, on the other. Another well-established instance of the gauge/gravity duality is the pair constituted by M-Theory on $AdS_4\times S^7/Z_k$ and the large $N$ limit of the ABJM  theory, which was introduced in \cite{Aharony:2008ug}. We see that at low energy these two instances involve General Relativity on $AdS_5$ and $AdS_4$ as duals of strongly interacting field theories without gravity in $4$ and $3$ dimensions, respectively.

The reader should bear in mind that from now on shall consider Euclidean AdS only. In Poincar\'e coordinates, Euclidean AdS is the space ${\cal H}_{d+1}=\{(z,\vec{x})\mid z>0, \vec{x}\in \mathbb{R}^d\}$ with line element
\begin{equation}
ds^2\,=\,\frac{L^2}{z^2}(dz^2+ \delta_{ij}dx^i dx^j).
\label{standardmetric}
\end{equation}
The (conformal) boundary of ${\cal H}_{d+1}$ is at $z=0$.

The gauge/gravity duality, when it holds, is a precisely formulated realization of the holographic principle \cite{tHooft:1993dmi, Susskind:1994vu}. The formulation in question entails the so-called holographic dictionary introduced in \cite{Witten:1998qj, Gubser:1998bc}. This dictionary sets a correspondence between objects (parameters and fields) of the quantum gravity theory in $d+1$-dimensions and the dual quantum field theory in $d$-dimensions. In particular, the quantum fluctuations, say $h_{\mu\nu}$, of the Euclidean AdS metric  is linked with the energy-momentum tensor, $T_{ij}$, of the dual quantum field theory. Indeed,  the data, say $h^{(b)}_{ij}$, setting the  value of $h_{\mu\nu}$ at the conformal boundary of Euclidean AdS acts a source of the energy-momentum tensor of the dual quantum field theory: it is postulated that the $n$-point connected Green function of $T_{ij}$  is given by
\begin{equation}
\langle T_{i_1 j_1}(x_1)\cdots T_{i_n j_n}(x_n)\rangle^{(connected)}\;=\;\left.\frac{\delta^n\,Ln\,Z_{gravity}[h^{(b)}_{ij}]}{\delta h^{(b)i_1 j_1}(x_1)\cdots \delta h^{(b)i_n j_n}(x_n)}\right\vert_{h^{(b)}=0},
\label{EMcorr}
\end{equation}
where $Z_{gravity}[h^{(b)}_{ij}]$ is the partition function of the gravity theory on the Euclidean AdS background for the boundary data $h^{(b)}_{ij}$.

In this paper we shall be concerned only with the leading saddle point approximation to $Z_{gravity}[h_{ij}^{(b)}]$. This approximation is given by
 \begin{equation}
\ln\;Z_{gravity}[h^{(b)}_{ij}]\,=\,-S_{classical}[h_{\mu\nu}[h^{(b)}_{ij}]],
\label{saddlep}
\end{equation}
where $h_{\mu\nu}[h^{(b)}_{ij}]$ is the solution to the classical gravity equations of motion in the Euclidean AdS background with boundary data equal to $h^{(b)}_{ij}$.

Of course, as they stand, both (\ref{EMcorr}) and (\ref{saddlep}) are formal equations: they need regularization and renormalization to be well-defined. We shall regularize and renormalize $S_{classical}[h_{\mu\nu}[h^{(b)}_{ij}]]$ as done in \cite{Witten:1998qj, Freedman:1998tz, Liu:1998bu, Arutyunov:1998ve, DHoker:2002nbb}, ie, first, by cutting off at $\epsilon_0 >0$ the ``$z$" coordinate of the Euclidean AdS metric in  Poincar\'e coordinates; and, then, subtracting the divergences which arise as $\epsilon_0$ goes to zero. We shall not use the holographic renormalization framework of \cite{Henningson:1998gx} --see \cite{Skenderis:2002wp}, for a pedagogical exposition. This framework demands the use of the Graham-Fefferman form of the near boundary metric, which in not a unimodular metric.

The purpose of this paper to work out, in the leading saddle point approximation, the 2-point and 3-point contributions to the partition function --see (\ref{saddlep})-- of unimodular gravity for an Euclidean AdS background and thus to begin the analysis of the properties of unimodular gravity  from the gauge/gravity duality standpoint.
By unimodular gravity we shall mean a  gravity theory  as formulated by using the framework of references \cite{Alvarez:2006uu, Alvarez:2005iy, Alvarez:2015sba}. In the framework
in question the  unimodular metric, say $\hat{g}_{\mu\nu}$, is expressed in terms of the unimodular background metric $\bar{g}_{\mu\nu}$ and the unconstrained field $h_{\mu\nu}$ as follows
\begin{equation}
\hat{g}_{\mu\nu}=\frac{g_{\mu\nu}}{|g|^{1/n}}\,\quad g_{\mu\nu}=\bar{g}_{\mu\nu}+\kappa h_{\mu\nu}.
\label{gravitonfield}
\end{equation}
In the previous equations $g$ denotes the determinant of $g_{\mu\nu}$, $n$ is the space-time dimension and $\kappa=\sqrt{8\pi G}$; $G$ being the gravitational constant. The two-tensor $h_{\mu\nu}$ describes the perturbations of the background $\bar{g}_{\mu\nu}$, classically, and the fluctuations of the latter at the quantum level. Upon quantization $h_{\mu\nu}$ becomes the graviton field \cite{Alvarez:2006uu, Alvarez:2005iy}. The gauge symmetry of this formulation of unimodular gravity is constituted by transverse diffeomorphisms and  Weyl transformations of $g_{\mu\nu}$ \cite{Alvarez:2006uu, Alvarez:2016lbz}.

The classical action of our unimodular gravity theory for a manifold ${\cal M}$ with boundary $\partial{\cal M}$ is  \cite{Blas:2008uz, Fiol:2008vk}
\begin{equation}
 S_{\text{\tiny{UG}}}\,=\, -\frac{1}{2\kappa^2}\Big(\int_{\cal M}d^n x\,R[\hat{g}_{\mu\nu}]\,+\,2\int_{\partial\cal M}d^{n-1} y\,\sqrt{\hat{g}^{(b)}}K\Big),
 \label{UGaction}
\end{equation}
where $R[\hat{g}]$ is the Ricci scalar,  $\hat{g}_{(b)}$ is the determinant of the induced metric on the boundary and $K$ is the trace of the extrinsic curvature of the boundary for the unimodular metric $\hat{g}_{\mu\nu}$. Of course,$\hat{g}_{\mu\nu}$ is given in (\ref{gravitonfield}). The equation of motion derived from $S_{\text{\tiny{UG}}}$ reads  \cite{Alvarez:2015sba}
\begin{equation}
R_{\m\n}-\frac{1}{n}R g_{\m\n}={(n-2)(2n-1)\over 4 n^2} \left({\nabla_\m g\nabla_\n g \over g^2}-{1\over n} {(\nabla g)^2\over g^2} g_{\m\n}\right)-{n-2\over 2n} \left({\nabla_\m \nabla_\n g \over g}-{1\over n}
{\nabla^2 g\over g} g_{\m\n}\right),
\label{UGeom}
\end{equation}
where $R_{\mu\nu}$ and $R$ are the Ricci tensor and the Ricci scalar for $g_{\mu\nu}$ --not for $\hat{g}_{\mu\nu}$, respectively; $\nabla_\mu g\equiv \partial_\mu g$. The previous equations, which we shall call the unimodular equation of motion,  are obtained by setting to zero the infinitesimal  variations of $S_{\text{\tiny{UG}}}$ induced by infinitesimal variations of $g_{\mu\nu}$ which vanish at $\partial{\cal M}$.

The reader should notice that no Cosmological Constant occurs in $S_{\text{\tiny{UG}}}$ and yet $g_{\mu\nu}=\bar{g}_{\mu\nu}$ is a solution to the unimodular equation of motion  in (\ref{UGeom}) when $\bar{g}_{\mu\nu}$ is the unimodular Euclidean AdS  metric. This result holds whatever the value of the Cosmological Constant which occurs in the  Euclidean AdS metric. This is in sharp contrast with the General Relativity situation where the Cosmological Constant enters the action and the value of Cosmological Constant which characterizes the Euclidean AdS metric is only the one which occurs in the action.

We shall show that the two-  and three-point contributions to the r.h.s of (\ref{saddlep}) in General Relativity and unimodular gravity are not the same for the IR regularized theories. However, this difference is due only to IR divergent contact contributions so that once these IR divergent terms are  subtracted full agreement between the unimodular gravity and General Relativity results is reached. As a consequence, the two-point and three-point correlation functions of the energy momentum tensor defined according to (\ref{EMcorr}) are the same for both  gravity theories. And yet, this equivalence between unimodular gravity and General Relativity regarding those IR finite results cannot hide the fact that it is obtained in a non trivial way.

The layout of this paper is as follows. In section 2  we  put forward  the unimodular counterpart of Euclidean AdS in  Poincar\'e coordinates. In section 3 we solve the linearized version of unimodular gravity equation (\ref{UGeom}) for  the unimodular Euclidean AdS background. We shall show that a suitable gauge choice --the axial gauge-- and coordinates turns the linearized equation in question into the equation of a free massless scalar field on the Euclidean AdS background. Sections 4 and 5 are devoted, respectively, to the computation of the two-  and three-point contributions to the r.h.s. of (\ref{saddlep}) for unimodular gravity and how these contributions compare to their General Relativity counterparts. In section 6 we shall state our conclusions. We also include an Appendix where we discuss how to find the solution to the linearized General relativity equations in the axial gauge, the solution satisfying Dirichlet Boundary conditions and having a well-defined limit as we move towards the interior of Euclidean AdS.

\section{Euclidean AdS with unimodular metric. Unimodular Poincar\'e coordinates.}

In the standard Gauge/Gravity duality discussions \cite{DHoker:2002nbb}, one usually characterises Euclidean AdS by using Poincar\'e coordinates, and thus Euclidean AdS in $d+1$ dimensions is identified with the set of $\rm{I\!R}^{d+1}$ points $\{(z,\vec{x}), z > 0, \vec{x}\in \rm{I\!R}^{d}\}$ with line element
\begin{equation}
ds^2\,=\,\frac{L^2}{z^2}\left(dz^2+\delta_{ij}dx^i dx^j\right),\quad i,j=1\ldots d.
\label{linepoincare}
\end{equation}
In this coordinate system the boundary is at $z=0$ and it is $\rm{ I\!R}^{d}$.

The determinant of the metric of the previous line element is not $1$, so this metric does not suit our purposes. Let us introduce a new coordinate, say $w$, $w\geq 0$, defined as follows
\begin{equation}
w\,=\,\frac{L^{d+1}}{d}z^{-d}.
\label{wcoord}
\end{equation}
Here and elsewhere $d\geq 3$. In terms of $w$ the line element in (\ref{linepoincare}) reads
\begin{equation}
ds^2\,=\,\left(\frac{L}{w d}\right)^2 dw^2+\left(\frac{w d}{L}\right)^{2/d}\delta_{ij}dx^i dx^j.
\label{UGline}
\end{equation}
The Riemannian metric of the line element in (\ref{UGline}) is unimodular; but now Euclidean AdS is identified with  set of real $d\!+\!1$-tuples $(w,\vec{x})$, $w > 0$, $\vec{x}\in \rm{I\!R}^{d}$  and the boundary is at $w=\infty$.

The graviton field, $h_{\mu\nu}$, of our unimodular gravity theory will propagate in an Euclidean AdS  background with unimodular metric $\bar{g}_{\mu\nu}$ --the background metric-- given by
\begin{equation}
\bar{g}_{\mu\nu}(w,\vec{x})\,=\,\left(\left(\frac{L}{w d}\right)^2,\left(\frac{w d}{L}\right)^{2/d}\delta_{ij}\right),
\label{unimetric}
\end{equation}
where $\mu,\nu=0,1...d$ and $i,j=1...d$.

Let us close this section by making some comments regarding the killing vectors of a general unimodular metric. First, any such killing vector, $\xi^{\mu}$, is transverse, ie, $\partial_{\mu}\xi^\mu=0$, since transversality is equivalent to covariant transversality, $\nabla_\mu \xi^\mu=0$, when the metric is unimodular. Secondly, the number of independent killing vectors of a unimodular metric and any metric obtained from it by a diffeomorphism is the same. This is relevant with regard to the gauge/gravity duality.\footnote{We thank E. \'Alvarez for pointing out these two results to us.}

\section{The linearized unimodular gravity equation on an Euclidean AdS background.}

The linearized unimodular gravity equation in the Euclidean AdS background with the unimodular metric, $\bar{g}_{\mu\nu}$, in (\ref{unimetric}) is obtained from the equation in (\ref{UGeom}) with $n=d+1$,  by setting $g_{\mu\nu}=\bar{g}_{\mu\nu}+\kappa h_{\mu\nu}$ and expanding at first order in $\kappa$. Thus, one gets
\begin{equation}
\begin{array}{l}
{\frac{1}{2}\bar{\Box}h_{\mu\nu}-\frac{d+3}{2(d+1)^2}\bar{g}_{\mu\nu}\bar{\Box}h-\frac{1}{2}\bar{\nabla}_\mu\bar{\nabla}_\rho h^\rho_\nu-\frac{1}{2}\bar{\nabla}_\nu\bar{\nabla}_\rho h^\rho_\mu+\frac{1}{d+1}\bar{g}_{\mu\nu}\bar{\nabla}_\rho\bar{\nabla}_\sigma h^{\rho\sigma}
+\frac{1}{d+1}\bar{\nabla}_\mu\bar{\nabla}_\nu h}\\[4pt]
{+\frac{1}{L^2}h_{\mu\nu}-\bar{g}_{\mu\nu}\frac{1}{(d+1)L^2} h=0,}
\end{array}
\label{linearisedugeq}
\end{equation}
where all the covariant derivatives are defined with respect to $\bar{g}_{\mu\nu}$ --hence, the upper bar-- and $h\equiv\bar{g}^{\mu\nu}\,h_{\mu\nu}$. Let us point out that (\ref{linearisedugeq}) is quite different from the corresponding General Relativity equation, (\ref{linGR}), in the Appendix.

The aim of this section is to find the solution to (\ref{linearisedugeq}) for suitable  Dirichlet data at the boundary and such that --see \cite{DHoker:2002nbb, Ammon:2015wua}-- the solution in question has a well-defined limit as one moves deep into the interior of Euclidean AdS, ie, as $w\rightarrow 0$. We shall cut-off  the $w$ coordinate at $\rho_{0}$ --ie, $0\leq w\leq \rho_{0}$--  to regularize the IR divergent contributions to the r.h.s of (\ref{saddlep}) coming from regions arbitrarily close to $w=\infty$. Thus, we shall solve  (\ref{linearisedugeq}) in the domain $\{(w,\vec{x}); 0 < w < \rho_{0}, \vec{x}\in \rm{I\!R}^d\}$. We shall show that in the axial gauge,
$h_{0\mu}[w,\vec{x}]=0$, such a solution can be brought to a solution, say $h_{\mu\nu}=(h_{0\mu}=0,h_{ij})$, satisfying
\begin{equation}
\delta^{ij}h_{ij}[w,\vec{x}]\,=\,0\quad\text{and}\quad\partial^j h_{ji}[w,\vec{x}]\,=\,0,
\label{TTpremier}
\end{equation}
by doing a gauge transformation that preserves the axial gauge condition. In (\ref{TTpremier}), $i,j=1...d$ and
$\partial^j=\delta^{jl}\frac{\partial}{\partial x^l}$.

To solve (\ref{linearisedugeq}) for $h_{\mu\nu}$, we shall take advantage of the gauge symmetries:
\begin{equation}
\begin{array}{l}
{\delta h_{\mu\nu}(x)\,=\,\bar{\nabla}_\mu\theta_\nu(x)+\bar{\nabla}_\nu\theta_\mu(x),\quad \bar{\nabla}_\mu\theta^\mu(x)=0,}\\[4pt]
{\delta_{W} h_{\mu\nu}(x)(x)= 2\sigma(x)\bar{g}_{\mu\nu},\quad x\equiv (w,\vec{x}).}
\label{gaugesym}
\end{array}
\end{equation}
of the equation in question. $\bar{\nabla}_\mu$ is defined with regard to the unimodular metric $\bar{g}_{\mu\nu}$ in (\ref{unimetric}). That the transformations in (\ref{gaugesym}) leave (\ref{linearisedugeq}) invariant can be easily checked directly and it is a consequence of the fact  --see \cite{Alvarez:2015sba}-- that the unimodular action in (\ref{UGaction}) is invariant under transverse diffeomorphisms and Weyl transformations  of $g_{\mu\nu}$ in (\ref{gravitonfield}). Recall that $\partial_\mu\theta^\mu=0$ is equivalent to $\nabla_\mu\theta^\mu(x)=0$ if the metric is unimodular.

By using the transformations in (\ref{gaugesym}),  one may impose the gauge condition $h_{0\mu}[w,\vec{x}]=0$, $0\leq w\leq \rho_0$ and $\vec{x}\in\rm{I\!R}^d$.  From now on we shall assume that the previous gauge condition is imposed so that only $h_{ij}[w,\vec{x}]$ occurs in (\ref{linearisedugeq}).

Let us introduce the following definitions
\begin{equation*}
\begin{array}{l}
{H_{ij}[z,\vec{x}]\,=\,h_{ij}[w=\frac{L^{d+1}}{d}z^{-d},\vec{x}],\quad H[z,\vec{x}]=\delta^{ij}H_{ij}[z,\vec{x}],\quad i,j=1...d}\\[4pt]
{f[z,\vec{x}]\,=\,\idkd\;f[z,\vec{k}]\;e^{-i\vec{k}\cdot\vec{x}},\quad\vec{k}=(k^1,...,k^d),\quad f''=\frac{d^2 f}{dz^2},\quad f'=\frac{d f}{dz}.}
\end{array}
\end{equation*}
Then, after changing variables from $w$ to $z=(\frac{w d}{L^{d+1}})^{-1/d}$ equation (\ref{linearisedugeq}) boils down to the following set of equations
\begin{equation}
\begin{array}{l}
{H''[z,\vec{k}]((-1 + d) z^2)+H'[z,\vec{k}](-((-5 + d) (-1 + d) z))+}\\[4pt]
{H[z,\vec{k}](-2 (-2 + d) (-1 + d) + (3 + d) k^2 z^2)-k^i k^j H_{ij}[z,\vec{k}](2 (1 + d) z^2)=0,}
\end{array}
\label{eom00}
\end{equation}
\begin{equation}
-2 k_i z H'[z,\vec{k}]+(-5+d)k_iH[z,\vec{k}]+ (1+d)( z k^j H_{ji}'[z,\vec{k}]+2 k^j H_{ji}[z,\vec{k}])=0.
\label{eom0i}
\end{equation}
\begin{equation}
\begin{array}{l}
{H''[z,\vec{k}] (-(3 + d) z^2)\delta_{ij}+ H'[z,\vec{k}]((-5 + d) (3 + d) z\delta_{ij} )+}\\[8pt]
{ H[z,\vec{k}](-2 (1 + d) z^2 k_i k_j  + (3 + d)(-4 + 2 d + k^2 z^2)\delta_{ij})+}\\[8pt]
{H_{ij}''[z,\vec{k}](1 + d)^2 z^2+ H_{ij}'[z,\vec{k}](-(-5 + d) (1 + d)^2  z)+
H_{ij}[z,\vec{k}](-(1 + d)^2  (-4 + 2 d + k^2 z^2))+}\\[8pt]
{(- 2 (1 + d) z^2)\delta_{ij} k^lk^m H_{lm}[z,\vec{k}] +
  (1 + d)^2 z^2 (k_j k^l H_{li}[z,\vec{k}] + k_i k^l H_{lj}[z,\vec{k}])=0.}
\end{array}
\label{eomij}
\end{equation}
Let us stress that equations (\ref{eom00}), (\ref{eom0i}) and (\ref{eomij}) are equivalent to the components $00$, $0i$ and $ij$ of equation (\ref{linearisedugeq}), respectively. $i,j$ run from $1$ to $d$.

Let us first show that (\ref{eom00}), (\ref{eom0i}) and (\ref{eomij}) imply that, modulo a transverse diffeomeorphism transformation that preserves $h_{0\mu}[w,\vec{x}]=0$,
\begin{equation}
H[z,\vec{k}]=0\quad \text{and}\quad k^jH_{ji}[z,\vec{k}]=0,
\label{TTconditions}
\end{equation}
when $h_{ij}[w,\vec{x}]$ has a well-defined limit as $w\rightarrow 0$.
To do this we shall proceed as follows. Contracting equation (\ref{eom0i}) with $k^i$ one gets
\begin{equation}
{ -2 k^2 z H'[z,\vec{k}]+ (-5+d) k^2 H[z,\vec{k}] + (1 + d) z k^i k^j H_{ij}'[z,\vec{k}]+2(1+d) k^i k^j H_{ij}[z,\vec{k}] =0.}
\label{kieom0i}
\end{equation}
By taking the derivative with respect to $z$ of the previous equation, one obtains
\begin{equation}
{(1+d) z k^i k^j H_{ij}''[z,\vec{k}]+ 3 (1+d)k^i k^j H_{ij}'[z,\vec{k}] - 2 k^2 z H''[z,\vec{k}]+ (-7 + d) k^2 H'[z,\vec{k}]=0.}
\label{derivativekieom0i}
\end{equation}
Let us now contract (\ref{eomij}) with $k^i k^j$:
\begin{equation}
\begin{array}{l}
{H''[z,\vec{k}] (-(3 + d) k^2 z^2 )+ H'[z,\vec{k}](-(-5 + d) (3 + d) k^2 z)+}\\[8pt]
{ H[z,\vec{k}](2 (-6 + d + d^2) k^2 - (-1 + d) k^4 z^2 )+}\\[8pt]
{k^i k^jH_{ij}''[z,\vec{k}]((1 + d)^2 z^2)+ k^i k^jH_{ij}'[z,\vec{k}](-(-5 + d) (1 + d)^2 z )+}
\\[8pt]
{k^i k^jH_{ij}[z,\vec{k}](4+6d-2d^3 + (-1 + d^2) k^2 z^2)=0.}
\end{array}
\label{kikjeomij}
\end{equation}

Let us consider the system constituted by (\ref{kieom0i}), (\ref{derivativekieom0i}) and (\ref{kikjeomij}). Solving this system for $k^i k^j H_{ij}[z,\vec{k}]$, one gets
\begin{equation}
\begin{array}{l}
{k^i k^j H_{ij}[z,\vec{k}]=\frac{1}{(1 + d) z^2}\{(-2 - (-3 + d) d + k^2 z^2) H[z,\vec{k}] +
 z (2 (-2 + d) H'[z,\vec{k}] - z H''[z,\vec{k}])\} .}
\end{array}
\label{eq1}
\end{equation}
The contraction of equation (\ref{eomij}) with $\delta^{ij}$ yields the following equation
\begin{equation}
\begin{array}{l}
{H''[z,\vec{k}] (-(-1+d)z^2)+
H'[z,\vec{k}]((-5 + d) (-1 + d) z)+}\\[8pt]
{H[z,\vec{k}](2 (-2 + d) (-1 + d) - (3 + d) k^2 z^2  )+
k^ik^jH_{ij}[z,\vec{k}](2 (1 + d) z^2)=0.}
\end{array}
\label{preveq}
\end{equation}
This is equation (\ref{eom00}), so we conclude that equation (\ref{eom00}) is contained in equation (\ref{eomij}) and provides no extra information.
By solving for $k^ik^jH_{ij}[z,\vec{k}]$, (\ref{preveq}) can be recast into the form
\begin{equation}
\begin{array}{l}
{k^ik^jH_{ij}[z,\vec{k}]=\frac{1}{2 (1 + d) z^2}\{(-2 (-2 + d) (-1 + d) + (3 + d) k^2 z^2) H[z,\vec{k}] + }\\[8pt]
{(-1 + d) z (-(-5 + d) H'[z,\vec{k}] + z H''[z,\vec{k}]\}.}
\end{array}
\label{eq2}
\end{equation}

Next, subtracting (\ref{eq2}) from (\ref{eq1}), one gets
\begin{equation*}
k^2 z H[z,\vec{k}] - (-3 + d) H'[z,\vec{k}] + z H''[z,\vec{k}]=0.
\end{equation*}
Since $k^2\geq 0$, the general solution to the previous equation reads
\begin{equation}
H[z,\vec{k}]=z^{-1+d/2}(C_1 J_{d/2-1}[ |\vec{k}|\, z]  + C_2 Y_{d/2-1}[ |\vec{k}|\, z]),
\label{badhz}
\end{equation}
where $|\vec{k}|=\sqrt{k^2}$, and $C_ 1$ and $C_2$ are  functions of $\vec{k}$.

Let us assume that $d\geq 3$. Then, the asymptotic behaviour of $J_{d/2-1}[|\vec{k}| z]$ and $Y_{d/2-1}[ |\vec{k}| z])$ leads to the conclusion that $H[z,\vec{k}]=\delta^{ij}H_{ij}[z,\vec{k}]$ in (\ref{badhz}) has a well-defined limit as $z\rightarrow\infty$  only if both $C_ 1$ and $C_2$ vanish. Recall that there is the condition that $H_{ij}[z,\vec{k}]=h_{ij}[w=\frac{L^{d+1}}{d}z^{-d},\vec{k}]$ must have a well-defined limit as $w\rightarrow 0$, ie, as $z\rightarrow\infty$.

 Next, the substitution of $H[z,\vec{k}]=0$ in equation (\ref{eom0i}) leads to
\begin{equation*}
z k^j H_{ji}'[z,\vec{k}]+2 k^j H_{ji}[z,\vec{k}]=0,
\end{equation*}
whose general solution is
\begin{equation*}
 k^i H_{ij}[z,\vec{k}]=\frac{v_j(\vec{k})}{z^2}.
\end{equation*}
This solution is compatible with equation (\ref{eom00})  for $H[z,\vec{k}]=0$ if, and only if,
\begin{equation}
 \delta^{ij}k_i v_j(\vec{k})=0.
\label{transverse}
\end{equation}

It can be shown that
\begin{equation}
H^{(particular)}_{ij}[z,\vec{k}]= \frac{1}{(z^2 k^2)}(k_i v_j (\vec{k})+ k_j v_i (\vec{k})).
\label{hparticular}
\end{equation}
is a solution to equation (\ref{eomij}), for $\delta^{ij}H^{(particular)}_{ij}[z,\vec{k}]=0$. Hence, when $H[z,\vec{k}]=0$, the general solution, $H_{ij}[z,\vec{k}]$, to (\ref{eomij}) can be expressed as the sum
$H_{ij}[z,\vec{k}]=H^{(transverse)}_{ij}[z,\vec{k}]+H^{(particular)}_{ij}[z,\vec{k}]$, where
\begin{equation*}
k^i H^{transverse}_{ij}[z,\vec{k}]=0.
\end{equation*}

Let us show that
\begin{equation*}
H^{(particular)}_{ij}[z,\vec{x}]\,=\,\idkd\;H^{(particular)}_{ij}[z,\vec{k}]\;e^{-i\vec{k}\cdot\vec{x}},
\end{equation*}
with $z=\left(\frac{w d}{L^{d+1}}\right)^{-1/d}$, can be recast as a unimodular gauge transformation which preserves the axial gauge condition $h_{0\mu}[w,\vec{x}]=0$. This gauge transformation reads
\begin{equation}
\nabla_{\mu}{\cal W}_{\nu}[w,\vec{x}]+\nabla_{\nu}{\cal W}_{\mu}[w,\vec{x}],
\label{Wtrans}
\end{equation}
where
\begin{equation}
{\cal W}_0[w,\vec{x}]=0,\quad {\cal W}_i[w,\vec{x}]= \Big(\frac{d\, w}{L^{d+1}}\Big)^{2/d}\,
i\int\frac{d^d k}{(2\pi)^d}\;e^{-i k\cdot x} \frac{v_i (\vec{k})}{k^2}
\label{Wdefinition}
\end{equation}
and the covariant derivative is defined with regard to the unimodular Poincar\'e metric in (\ref{unimetric}).

Let us change variables from $(w,\vec{x})$ to $(z,\vec{x})$, where $z= \left(\frac{w d}{L^{d+1}}\right)^{-1/d}$ --$\vec{x}$ does not change. Then the vector field ${\cal W}_{\nu}[w,\vec{x}]$ changes to ${\cal V}_{\nu}[x,\vec{x}]$ as follows
\begin{equation*}
{\cal W}_0[w,\vec{x}]=\frac{\partial z}{\partial w}\;{\cal V}_0[z,\vec{x}],\quad {\cal }W_{i}[w,\vec{x}]={\cal V}_i[z,\vec{x}].
\end{equation*}
Hence, the following results hold
\begin{equation}
\begin{array}{l}
{{\cal V}_0[z,\vec{x}]=0,\quad {\cal V}_i [z,\vec{x}]=\frac{1}{z^2}i\int\frac{d^d k}{(2\pi)^d}\;e^{-i k\cdot x} \frac{v_i (\vec{k})}{k^2}} \\[8pt]
{\bar{\nabla}_{0}{\cal W}_{0}[z,\vec{x}]=(\frac{\partial z}{\partial w})^2\nabla^{(S)}_{0}{\cal V}_{0}[z,\vec{x}],}\\[8pt]
{\bar{\nabla}_{0}{\cal W}_{i}[w,\vec{x}]+\bar{\nabla}_{i}{\cal W}_{0}[w,\vec{x}]=\frac{\partial z}{\partial w}(\nabla^{(S)}_{0}{\cal V}_{i}[z,\vec{x}]+\nabla^{(S)}_{i}{\cal V}_{0}[z,\vec{x}]),}\\[8pt]
{\bar{\nabla}_{i}{\cal W}_{j}[w,\vec{x}]+\bar{\nabla}_{j}{\cal W}_{i}[w,\vec{x}]=\nabla^{(S)}_{i}{\cal V}_{j}[z,\vec{x}]+\nabla^{(S)}_{j}{\cal V}_{i}[z,\vec{x}],}
\end{array}
\label{someres}
\end{equation}
where $\nabla^{(S)}_{\mu}$ denotes the covariant derivative with respect to the standard Poincar\'e metric whose line element is in (\ref{linepoincare}). A little computation yields
\begin{equation*}
\nabla^{(S)}_{0}{\cal V}_{0}[z,\vec{x}]=0,\quad \nabla^{(S)}_{0}{\cal V}_{i}[z,\vec{x}]+\nabla^{(S)}_{i}{\cal V}_{0}[z,\vec{x}]=0,
\end{equation*}
which guarantees, in view of (\ref{gaugesym}) and (\ref{someres}), that the axial gauge condition $h_{0\mu}[w,\vec{x}]=0$ is preserved. Besides
\begin{equation*}
\nabla^{(S)}_{i}{\cal V}_j[z,\vec{x}]+\nabla^{(S)}_{j}{\cal V}_{i}[z,\vec{x}]=\frac{1}{z^2}\int\frac{d^d k}{(2\pi)^d}\;e^{-i k\cdot x}\; \frac{k_i v_j (k)+ k_j v_i(k)}{k^2},
\end{equation*}
which matches (\ref{hparticular}). Hence, the last equation in (\ref{someres}) yields (\ref{hparticular}).

It remains to be seen that ${\cal W}_{\mu}[w,\vec{x}]$ is covariantly transverse: $\bar{\nabla}^{\mu}{\cal W}_{\mu}[w,\vec{x}]=0$. Indeed,
\begin{equation*}
\bar{\nabla}^{\mu}{\cal W}_{\mu}[w,\vec{x}]=\nabla^{(S)}_{\mu}{\cal V}^{\mu}[z,\vec{x}]=\int\frac{d^d k}{(2\pi)^d}\;e^{-i k\cdot x}\; \frac{\delta^{ij} k_i v_j (k)}{k^2} =0,
\end{equation*}
for equation (\ref{transverse}) holds. Recall that unimodularity of the metric implies that transversality with regard to $\partial_\mu$ and $\bar{\nabla}_\mu$ are equivalent.

Let us recapitulate. We have just shown that, in the axial gauge, $h_{0\mu}[w,\vec{x}]=0$, any solution to (\ref{linearisedugeq}) in the  domain with cutoff $\{(w,\vec{x}), 0 < w<\rho_0, \vec{x}\in \rm{I\!R}^d\}$ which has a well-defined limit as $w\rightarrow 0$ is gauge equivalent, under the transformation in (\ref{Wtrans}) and (\ref{Wdefinition}), to a solution of (\ref{linearisedugeq}), say $h_{ij}[z,\vec{x}]$, such that
\begin{equation}
H[z,\vec{x}]\,=\,0\quad\text{and}\quad \partial^j H_{ji}[z,\vec{x}]=0,
\label{TTshown}
\end{equation}
where $\partial^j=\delta^{jl}\frac{\partial}{\partial x^l}$, $H[z,\vec{x}]\equiv\delta^{ij}H_{ij}[z,\vec{x}]$ and $H_{ij}[z,\vec{x}]=h_{ij}[w=\frac{L^{d+1}}{d}z^{-d},\vec{x}]$. Notice that (\ref{TTshown}) can be recast as (\ref{TTpremier}).

If we substitute (\ref{TTconditions}) in (\ref{eom00}) and (\ref{eom0i}) in turn, we shall see that they are trivially satisfied. However, the  substitution of (\ref{TTconditions}) in (\ref{eomij}) yields the following equation
\begin{equation}
z^2 H_{ij}''[z,\vec{k}]-(-5+d) z H_{ij}'[z,\vec{k}]-(-4 + 2 d + k^2 z^2) H_{ij}[z,\vec{k}]\,=\,0,
\label{simpleeomij}
\end{equation}
to be satisfied by $H_{ij}[z,\vec{k}]$.
Let  $H^i_j[z,\vec{k}]$ be given by the following set of equations
\begin{equation*}
H^i_j[z,\vec{k}]\,\equiv\,h^i_j[w=\frac{L^{d+1}}{d}z^{-d},\vec{k}],\quad h^i_j[w,\vec{k}]=\bar{g}^{il}h_{lj}[w,\vec{k}]=
\left(\frac{L}{w d}\right)^{2/d} h_{ij}[w,\vec{k}],
\end{equation*}
where $\bar{g}^{\mu\nu}$ is the inverse of the unimodular metric (\ref{unimetric}). Obviously, $H_{ij}[z,\vec{k}]=\frac{L^2}{z^2}H^i_j[z,\vec{k}]$, which substituted in (\ref{simpleeomij}) yields
\begin{equation}
z^2 H^{i}_j {''}[z,\vec{k}]+(1-d) z H^{i}_j{'}[z,\vec{k}]-k^2 z^2 H^i_j[z,\vec{k}]\,=\,0.
\label{besseleomij}
\end{equation}
The general solution to this equation is well known: it is a linear combination of $z^{d/2}K_{d/2}[|k|z]$ and $z^{d/2}I_{d/2}[|k|z]$, where $K_{d/2}[|k|z]$ and $I_{d/2}[|k|z]$ are the modified Bessel function of second kind. And yet, we have to drop $z^{d/2}I_{d/2}[|k|z]$, for it has an exponentially divergent behaviour in the deep interior of Euclidean AdS, ie, as $z\rightarrow\infty$ --recall that $z\rightarrow\infty$  corresponds to $w\rightarrow 0$. We then conclude that the solution to (\ref{besseleomij}),   in the domain
$\{(z,\vec{k}), z\!>\!\epsilon_0\!>\!0, \vec{k}\in \rm{I\! R}^d, \epsilon_0=\left(\frac{\rho_0 d}{L^{d+1}}\right)^{-1/d}\}$,
satisfying Dirichlet boundary conditions at $z=\epsilon_0$ and having a well-defined limit as $z\rightarrow 0$ reads
\begin{equation}
H^i_j[z,\vec{k}]\,=\, \frac{z^{d/2}K_{d/2}[|k|z]}{\epsilon_0^{d/2}
K_{d/2}[|k|\epsilon_0]}\,h^{(\tiny{TT})\,i}_j[\vec{k}].
\label{HIJresult}
\end{equation}
Notice that $h^{(\tiny{TT})\,i}_j[\vec{k}]$ is any traceless  and transverse function whose inverse Fourier transform is real so that (\ref{TTconditions}) holds.  Obviously,
\begin{equation}
\begin{array}{l}
{h^{(\tiny{TT})\,i}_j[\vec{k}]=h^{(T)\,i}_j[\vec{k}]-\frac{1}{d-1}\left(\delta^i_j-\frac{k^i k_j}{k^2}\right)h^{(T)}[\vec{k}], \quad h^{(T)}[\vec{k}]=\delta^j_i h^{(T)\,i}_j[\vec{k}], }\\[8pt]
{h^{(T)\,i}_j[\vec{k}]=h^{(b)\,i}_j[\vec{k}]-\frac{1}{k^2}k^i k_l h^{(b)\,l}_j[\vec{k}]-\frac{1}{k^2}k_j k^l h^{(b)\,i}_{l}[\vec{k}]+\frac{1}{(k^2)^2}k^i k_j\, k^n k_m h^{(b)\,m}_n[\vec{k}],}
\end{array}
\label{hTTdecomp}
\end{equation}
where $h^{(b)\,i}_j[\vec{k}]$ is the Fourier transform of an arbitrary real $h^{(b)\,i}_j(\vec{x})$, which sets the value of $h_{\mu\nu}[w,\vec{x}]$ at boundary   $w=\rho_{0}$.

Putting it all together we finally conclude that in the axial gauge, $h_{0\mu}[w,\vec{x}]=0$, any solution to (\ref{linearisedugeq}) --the linearized unimodular gravity equation-- in the domain $\{(w,\vec{x}); 0 < w < \rho_{0}, \vec{x}\in \rm{I\!R}^d\}$ is gauge equivalent,
under a gauge transformation --see (\ref{Wdefinition})-- preserving the axial gauge,  to an $h_{\mu\nu}[w,\vec{x}]$ whose Fourier transform is given by
\begin{equation}
\begin{array}{l}
{h_{0\mu}[w,\vec{k}]=0,\quad h_{ij}[w,\vec{k}]=\bar{g}_{ik}\,h^k_j[w,\vec{k}],}\\[8pt]
{h^k_j[w,\vec{k}]=H^k_j[z=\left(wd/L\right)^{-1/d},\vec{k}]=\displaystyle\left(\frac{\rho_0}{w}\right)^{1/2}\,\displaystyle\frac{K_{d/2}[|k|
\left(wd/L\right)^{-1/d}]}{K_{d/2}[|k|\left(\rho_0 d/L\right)^{-1/d}]}\,h^{(\tiny{TT})\,k}_j[\vec{k}].}
\end{array}
\label{finalresult}
\end{equation}
Of course, we have demanded that the solution, $h_{\mu\nu}[w,\vec{x}]$, be such that it has a well-defined limit as $w\rightarrow 0$ and satisfies  Dirichlet boundary conditions at $w=\rho_0$.

It will be useful for use in the following sections to realize that in the axial gauge, $h_{0\mu}[w,\vec{x}]=0$, the  equations in (\ref{TTpremier}) are equivalent to
\begin{equation}
h[w,\vec{x}]=\bar{g}^{\mu\nu}h_{\mu\nu}[w,\vec{x}]\quad \text{and}\quad \bar{\nabla}^\mu h_{\mu\nu}[w,\vec{x}]=0,
\label{TTcovariant}
\end{equation}
respectively; $\bar{g}_{\mu\nu}$ being defined in (\ref{unimetric}). Besides, the substitution of the equations  (\ref{TTcovariant}) in (\ref{linearisedugeq}) leads to the conclusion that our $h_{\mu\nu}[w,\vec{x}]$ in (\ref{finalresult}) satisfies
\begin{equation}
\bar{\Box}h_{\mu\nu}=-\frac{2}{L^2} h_{\mu\nu}.
\label{simpleeom}
\end{equation}

A final comment: It is not difficult to show that each component of $h^i_j[w,\vec{x}]$, with Fourier transform in (\ref{finalresult}), satisfies the free massless Klein-Gordon equation for the unimodular metric in (\ref{unimetric}).

\section{The two-point function}

The purpose of this section is to work out the expansion up to quadratic order  in $h_{\mu\nu}$ of $S_{\text{\tiny{UG}}}$ in (\ref{UGaction}) for the $h_{\mu\nu}$  in (\ref{finalresult}) and compare the result with that of General Relativity.

By using integration by parts and not dropping the total derivative terms, the contribution in question, say $S_{\text{\tiny{HEUG2}}}$, to
\begin{equation*}
 -\frac{1}{2\kappa^2}\int d^d x\int_{0}^{\rho_0}dw \,R[\hat{g}]
\end{equation*}
reads
\begin{equation}
\begin{array}{l}
{S_{\text{\tiny{HEUG2}}}[h_{\mu\nu}]=-\frac{1}{2\kappa^2}\int
d^d x\int_{0}^{\rho_0}dw \Big\{-\frac{d(d+1)}{L^2}+\kappa\bar{\nabla}_\mu\bar{\nabla}_\nu h^{\mu\nu}-\kappa\frac{1}{d+1}\bar{\Box} h+}\\[4pt]
{+\frac{\kappa^2}{2}\big[\frac{1}{2}h^{\alpha\beta}\bar{\Box}h_{\alpha\beta}-\frac{d+3}{2(d+1)^2}h\bar{\Box}h-\frac{1}{2}h^{\alpha\beta}
\bar{\nabla}_\alpha \bar{\nabla}_\lambda h^\lambda_\beta-\frac{1}{2}h^{\alpha\beta}\bar{\nabla}_\beta\bar{\nabla}_\lambda h^\lambda_\alpha +}\\[4pt]
{+\frac{1}{d+1}h\bar{\nabla}_\mu\bar{\nabla}_\nu h^{\mu\nu}+\frac{1}{d+1}h^{\alpha\beta}\bar{\nabla}_\alpha\bar{\nabla}_\beta h-\frac{1}{L^2}h^{\alpha\beta}h_{\alpha\beta}+\frac{1}{(d+1)L^2} h^2\big]+\kappa^2\bar{\nabla}_\lambda B^\lambda\Big\},}
\end{array}
\label{HEUG2}
\end{equation}
where\footnote{To obtain (\ref{HEUG2}), we have used the algebraic package xAct \cite{xAct}.}
\begin{equation*}
 B^\lambda=\frac{d-1}{4(d+1)^2} h\bar{\nabla}^\lambda h+\frac{3-d}{4(d+1)}h_{\mu\nu}\bar{\nabla}^\lambda h^{\mu\nu}+\frac{1}{2(d+1)}\Big[h^{\lambda\nu}\bar{\nabla}_\nu h+h\bar{\nabla}_\nu h^{\lambda\nu}\Big]-h^{\lambda\tau}\bar{\nabla}_\nu h^\nu_\tau-\frac{1}{2}h^{\tau\nu}\bar{\nabla}_\nu h^\lambda_\tau
 \end{equation*}
and $h\equiv \bar{g}^{\mu\nu}h_{\mu\nu}$. Notice that we are integrating over the domain with cutoff $\{(w,\vec{x}); 0\leq w\leq \rho_{0}, \vec{x}\in \rm{I\!R}^d\}$ that we have introduced in the previous section. The $\rm{I\!R}^d$ boundary is at $w=\rho_0$. The introduction of the cutoff $\rho_{0}$ regularizes the otherwise IR divergent value of the action. $\rho_0$ is to be taken to $\infty$ upon renormalization.

When $h_{\mu\nu}$ in (\ref{HEUG2}) satisfies --as does our solution in (\ref{finalresult})-- the equations in (\ref{TTcovariant}) and (\ref{simpleeom}), $S_{\text{\tiny{HEUG2}}}[h_{\mu\nu}]$ boils down to
\begin{equation}
S_{\text{\tiny{HEUG2}}}[h_{\mu\nu}]=-\frac{1}{2}\int
d^d x\int_{0}^{\rho_0}dw\,\left\{-\frac{d(d+1)}{\kappa^2 L^2}+\bar{\nabla}_\lambda\left(\frac{3-d}{4(d+1)}h_{\mu\nu}\bar{\nabla}^\lambda h^{\mu\nu}
-\frac{1}{2}h^{\tau\nu}\bar{\nabla}_\nu h^\lambda_\tau\right)\right\}.
\label{HEUG2shell}
\end{equation}
Notice that --as in the General Relativity case \cite{Liu:1998bu, Arutyunov:1998ve}-- $S_{\text{\tiny{HEUG2}}}$ in (\ref{HEUG2shell}) only contains boundary contributions.

Let us introduce the metric, say $\bar{g}^{(b)}_{ij}; i,j=1...d$, that the unimodular metric in (\ref{unimetric}) induces on the boundary, $\{(\rho_0,\vec{x}), \vec{x}\in \rm{I\!R}^d\}$, at $w=\rho_0$:
\begin{equation}
\bar{g}^{(b)}_{ij}[\rho_0,\vec{x}]=\bar{g}_{\mu\nu}[\rho_0,\vec{x}]\,\frac{\partial x^{\mu}}{\partial x^i}\frac{\partial x^{\nu}}{\partial x^j}= \bar{g}_{ij}[\rho_0,\vec{x}]= \left(\frac{\rho_0 d}{L}\right)^{2/d}\,\delta_{ij},
\label{bbgmetric}
\end{equation}
where $x^\mu=(w,x^i)$. Let $\bar{n}^{\mu}$ denote the unitary  vector which is orthogonal to the boundary $\{(\rho_0,\vec{x}), \vec{x}\in \rm{I\!R}^d\}$ and it is given by
\begin{equation}
\bar{n}^{\mu}=\left(\frac{\rho_0 d}{L},\vec{0}\right),
\label{nbarvalue}
\end{equation}
$\vec{0}$ being the zero vector of $\rm{I\!R}^d$. Of course, $\bar{n}^{\mu}$ satisfies $\bar{g}_{\mu\nu}\bar{n}^\mu \bar{n}^\nu=1$ and $\bar{g}_{\mu\nu}\bar{n}^\mu e^\nu_i=0$,  where $e^\mu_i=\frac{\partial x^{\mu}}{\partial x^i}, i=1...d$ are the coordinates of an orthogonal basis of the boundary at $w=\rho_0$ in the vector basis $\{\partial_\mu, \mu=0,1...d\}$. With this definitions in hand, the divergence theorem tell us that $S_{\text{\tiny{HEUG2}}}[h_{\mu\nu}]$
in (\ref{HEUG2shell}) is given by
\begin{equation}
S_{\text{\tiny{HEUG2}}}[h_{\mu\nu}]=-\frac{1}{2}\int
d^d x \left. \left\{-\frac{d(d+1)}{\kappa^2 L^2}\,w\,+\,\sqrt{\bar{g}^{(b)}}\,\bar{n}_\lambda\left(\frac{3-d}{4(d+1)}h_{\mu\nu}\bar{\nabla}^\lambda h^{\mu\nu}
-\frac{1}{2}h^{\tau\nu}\bar{\nabla}_\nu h^\lambda_\tau\right)\right\}\right\vert_{w=\rho_0},
\label{HEUG2shellgauss}
\end{equation}
where $\bar{g}^{(b)}$ denotes the determinant of $\bar{g}^{(b)}_{ij}$.

Now, substituting $h_{0\mu}=0$, $\bar{\nabla}_i h_{0j}=-\frac{1}{\omega d}h_{ij}$, $\sqrt{\bar{g}^{(b)}}=\frac{\rho_0 d}{L}$ and
(\ref{nbarvalue}) in (\ref{HEUG2shellgauss}), one gets
\begin{equation}
S_{\text{\tiny{HEUG2}}}[h_{\mu\nu}]=-\frac{1}{2}\int
d^d x \left. \left\{-\frac{d(d+1)}{\kappa^2 L^2}\,\rho_0\,+\,\left(\frac{\rho_0 d}{L}\right)^2\left(\frac{3-d}{4(d+1)}h^i_j\partial_0 h^j_i
+\frac{1}{2\rho_0 d}h^j_ih^i_j\right)\right\}\right\vert_{w=\rho_0}.
\label{HEUG2final}
\end{equation}

Next, we shall expand the unimodular Hawking-Gibbons-York action
\begin{equation}
 S_{\text{\tiny{HGY}}}\,=\, -\frac{1}{2\kappa^2}\int  d^dx\,2\sqrt{\hat{g}^{(b)}[\rho_0,\vec{x}]}\,K[\rho_0,\vec{x}],
 \label{HGYaction}
\end{equation}
up to second order in $h_{\mu\nu}$. Recall that $\hat{g}_{\mu\nu}$ is given in (\ref{gravitonfield}), with $n=d+1$, so that both the determinant of induced metric on the boundary, $g^{(b)}[\rho_0,\vec{x}]$, and the trace of the extrinsic curvature of the boundary, $K[\rho_0,\vec{x}]$, are to be computed for $\hat{g}_{\mu\nu}$.

Taking into account that
\begin{equation}
\hat{g}^{(b)}_{ij}[\rho_0,\vec{x}]\equiv\hat{g}_{\mu\nu}[\rho_0,\vec{x}]\,\frac{\partial x^{\mu}}{\partial x^i}\frac{\partial x^{\nu}}{\partial x^j}=\hat{g}_{ij}[\rho_0,\vec{x}],
\label{hgbij}
\end{equation}
where $x^\mu=(w,x^i)$, one concludes that in the axial gauge, $h_{0\mu}[\rho_0,\vec{x}]=0$, we have
\begin{equation}
\sqrt{\hat{g}^{(b)}}=\frac{\rho_0 d}{L}\Big[1+\frac{1}{2(d+1)}\kappa h-\frac{1}{4(d+1)}\kappa^2 h^i_jh^j_i+\frac{1}{8(d+1)^2}\kappa^2 h^2\Big]+o((h_{ij})^3),
\label{hgbdet}
\end{equation}
where $h=\bar{g}^{\mu\nu}h_{\mu\nu}$ and indices are raised  and lowered with the Euclidean AdS unimodular metric $\bar{g}_{\mu\nu}$ in (\ref{unimetric}).

To compute $K[\rho_0,\vec{x}]$ we shall take advantage of the foliation of $\{(w,\vec{x}); 0\leq w\leq \rho_{0}, \vec{x}\in \rm{I\!R}^d\}$ furnished
by the hyperplanes $w\times\rm{I\!R}^d$, $w$ fixed. Indeed, if $\hat{n}[w,\vec{x}]$ denotes the vector field constituted by the unitary vectors normal to each  hyperplane that we have just mentioned, we have
\begin{equation}
K[\rho_0,\vec{x}]\,=\,\hat{\nabla}_\mu n^{\mu}[\rho_0,\vec{x}]\,=\,\partial_{\mu} n^{\mu}[\rho_0,\vec{x}].
\label{Kcurvature}
\end{equation}
The covariant derivative $\hat{\nabla}_\mu$ is defined with regard to the metric $\hat{g}_{\mu\nu}$ which has determinant equal to $1$; this is why the rightmost equal sign in (\ref{Kcurvature}) is right. As we have said the vector field, $\hat{n}[w,\vec{x}]$, must satisfy the following unitarity and orthonormality conditions
\begin{equation}
\hat{g}_{\mu\nu} \hat{n}^\mu \hat{n}^\nu = 1\quad\text{ and}\quad \hat{g}_{\mu\nu}\hat{n}^\mu e^{\nu}_i=0,\quad i=1...d,
\label{unitarity}
\end{equation}
at each point $(w,\vec{x})$. In the previous equation $e^{\mu}_i=\frac{\partial x^\mu}{\partial x^i}$, $\{e^{\mu}_i \partial_{\mu}\}_{i=1...d}$ is a basis of vector fields of $w\times\rm{I\!R}^d$.

Let us solve the second equation in (\ref{unitarity}) first. Defining $\hat{n}_\mu=\hat{g}_{\mu\nu}\hat{n}^\nu$, we conclude that this second equation in (\ref{unitarity}) is equivalent to $\hat{n}_\mu e^\mu_i=0$. Hence,
\begin{equation}
\hat{n}_\mu[w,\vec{x}]\,=\,(n_{0}[w,\vec{x}],\vec{0}),
\label{nhat}
\end{equation}
for  $e^{\mu}_i=\frac{\partial x^\mu}{\partial x^i}=\delta^\mu_i$.

Now, in the axial gauge $h_{0\mu}=0$, so we have $\hat{g}_{i0}=0$, for $\bar{g}_{\mu\nu}$ is diagonal. Then, $\hat{n}_i=\hat{g}_{i\nu}\hat{n}^\nu=\hat{g}_{ij}\hat{n}^j$ and $\hat{n}_i=0$ imply that $(\bar{g}_{ij}+\kappa h_{ij})\hat{n}^j=0$; which in turn leads to
$\hat{n}^i[w,\vec{x}]=0$, for $(\bar{g}_{ij}+\kappa h_{ij})$ is an invertible matrix in perturbation theory of $h_{ij}$.

Summarizing, in the axial gauge, $h_{0\mu}=0$, the orthogonality condition -see (\ref{unitarity})-- on the vector field $\hat{n}^\mu$ yields
\begin{equation*}
\hat{n}^\mu[w,\vec{x}]\,=\,(n^{0}[w,\vec{x}],\vec{0}).
\end{equation*}
Substituting this result in the first equation --the unitarity condition-- in (\ref{unitarity}), one gets
\begin{equation*}
\hat{n}^0[w,\vec{x}]\,=\,\frac{1}{\sqrt{\hat{g}_{00}[w,\vec{x}]}}.
\end{equation*}

By taking into account that, in the axial gauge, it holds that $\hat{g}_{00}=\bar{g}_{00} (\text{det}(\bar{g}_{\mu\nu}+\kappa h_{\mu\nu})^{-1/(d+1)}$, one obtains the following result:
\begin{equation}
\hat{n}^0[w,\vec{x}]=\frac{ w d }{L}\Big[1+\frac{1}{2(d+1)}\kappa h-\frac{1}{4(d+1)}\kappa^2 h^i_j h^j_i+\frac{1}{8(d+1)^2}\kappa^2 h^2\Big]+o((h_{ij})^3).
\label{n0hat}
\end{equation}
The substitution of (\ref{nhat}) and (\ref{n0hat}) in (\ref{Kcurvature}) yields
\begin{equation}
\begin{array}{l}
{ K[\rho_0,\vec{x}]=\partial_\mu \hat{n}^\mu[\rho_0,\vec{x}]=\partial_0\hat{n}^0[w,\vec{x}]=\frac{d}{L}\left.\Big[1+\frac{1}{2(d+1)}\kappa h-\frac{1}{4(d+1)}\kappa^2 h^i_j h^j_i+\frac{1}{8(d+1)^2}\kappa^2h^2\Big]\right\vert_{w=\rho_0}+}\\[8pt]
{\phantom{K[\rho_0,\vec{x}]=}+\frac{\rho_0 d}{L}\left.\Big[\frac{1}{2(d+1)}\kappa \partial_0 h-\frac{1}{2(d+1)}\kappa^2 h^i_j\partial_0 h^j_i+\frac{1}{4(d+1)^2}\kappa^2 h\partial_0 h\Big]\right\vert_{w=\rho_0}+o((h_{ij})^3)}.
\end{array}
\label{Kresult}
\end{equation}
Notation: $\partial_0\equiv\frac{\partial}{\partial w}$.
Let us now substitute (\ref{hgbdet}) and (\ref{Kresult}) in (\ref{HGYaction}). Then,
\begin{equation}
\begin{array}{l}
{ S_{\text{\tiny{HGY}}}=-\frac{1}{2\kappa^2}\int d^dx\,2\Big[\frac{\rho_0 d^2}{L^2}\left(1+\frac{1}{d+1}\kappa h+\frac{1}{2(d+1)^2}\kappa^2 h^2-\frac{1}{2(d+1)}\kappa^2 h^i_jh^j_i\right)+}\\[8pt]
{+\left(\frac{\rho_0 d}{L}\right)^2\left(\frac{1}{2(d+1)}\kappa \partial_0 h+\frac{1}{2(d+1)^2}\kappa^2 h\partial_0 h-\frac{1}{2(d+1)}\kappa^2 h^i_j\partial_0 h^j_i\right)\left.\Big]\right\vert_{w=\rho_0}+o((h_{ij})^3).}
\end{array}
\label{preliminarySHGY}
\end{equation}
Recall that at the end of the day we have to replace $h_{\mu\nu}[w,\vec{x}]$ in the previous equation with the $h_{\mu\nu}[w,\vec{x}]$ in  (\ref{finalresult}). Then we can set $h=0$ in (\ref{preliminarySHGY}) to get
\begin{equation}
\begin{array}{l}
{ S_{\text{\tiny{HGY}}}=-\frac{1}{2\kappa^2}\int d^dx\,2\Big[\frac{\rho_0 d^2}{L^2}\left(1-\frac{1}{2(d+1)}\kappa^2 h^i_jh^j_i\right)
-\left(\frac{\rho_0 d}{L}\right)^2\left(\frac{1}{2(d+1)}\kappa^2 h^i_j\partial_0 h^j_i\right)\left.\Big]\right\vert_{w=\rho_0}+o((h_{ij})^3).}
\end{array}
\label{finalSHGY}
\end{equation}
To obtain the expansion of $S_{\text{\tiny{UG}}}$ in (\ref{UGaction}) up to second order in $h_{\mu\nu}$ for the solution in (\ref{finalresult}), all that is left for us to do is to add (\ref{HEUG2final}) and (\ref{finalSHGY}). Thus, we obtain
\begin{equation}
\begin{array}{l}
{S_{\text{\tiny{UG}}}\,=\,-\frac{1}{2\kappa^2}\times}\\[8pt]
{\int\!
d^d x\! \left. \left\{\frac{d(d-1)}{L^2}\rho_0\! +\!\kappa^2\frac{\rho_0}{L^2}\frac{d(1-d)}{2(d+1)} h^i_j[w,\vec{x}] h^j_i[w,\vec{x}]\!
-\!\kappa^2\left(\frac{\rho_0 d}{L}\right)^2\frac{1}{4} h^j_i[w,\vec{x}]\partial_0 h^i_j[w,\vec{x}]\right\}\right\vert_{w=\rho_0}\!\!+\!o((h_{ij})^3),}
\end{array}
\label{2ptaction}
\end{equation}
where $h^i_j[w,\vec{x}]$, or rather its Fourier transform, is given in (\ref{finalresult}).

To compare the result in (\ref{2ptaction}) with the corresponding results in General Relativity, which we shall borrow from \cite{Liu:1998bu} and  \cite{Arutyunov:1998ve}, we have to change coordinates from $(w,\vec{x})$ to $(z,\vec{x})$ by inverting the transformation in (\ref{wcoord}). Upon making this change of coordinates, one gets
\begin{equation}
\begin{array}{l}
{S_{\text{\tiny{UG}}}\,=\,-\frac{1}{2\kappa^2}\times}\\[8pt]
{\int\!
d^d x\! \left. \left\{\frac{(d-1)}{L}\big(\frac{\epsilon_0}{L}\big)^{-d}\!\!\!+\!\big(\frac{\epsilon_0}{L}\big)^{-d}\frac{\kappa^2(1-d)}{2(d+1)L}\! H^i_j[z,\vec{x}] H^j_i[w,\vec{x}]\!
+\!\left(\frac{\epsilon_0}{L}\right)^{1-d}\!\frac{\kappa^2}{4}\! H^j_i[z,\vec{x}]\partial_z H^i_j[z,\vec{x}]\right\}\right\vert_{z=\epsilon_0}\!\!\!\!+\!o((H_{ij})^3),}
\end{array}
\label{2ptactionz}
\end{equation}
where $\epsilon_0=(\rho_0 d/L^{d+1})^{-1/d}$ is the infrared cutoff for the $z$ variable. $H^i_j[z,\vec{x}]$ is defined its Fourier transform, which is  given in (\ref{HIJresult}) and (\ref{hTTdecomp}). $H^i_j[z,\vec{x}]$ occurs in (\ref{2ptactionz}) because of the definitions in (\ref{finalresult}). Notice that the second summand in (\ref{2ptactionz}) boils down to
\begin{equation*}
\Big(\frac{\epsilon_0}{L}\Big)^{-d}\,\frac{\kappa^2(1-d)}{2(d+1)L}\! h^{(\tiny{TT})\,i}_j[\vec{x}] h^{(\tiny{TT})\,j}_i[\vec{x}],
\end{equation*}
when $z$ is set to $\epsilon_0$. The Fourier transform of $h^{TT\,j}_i[\vec{x}]$ is given in (\ref{hTTdecomp}).

Now, $\epsilon_0$ is to be sent to $0$ (ie,$\rho_0\rightarrow\infty$) after subtracting the IR divergences regulated by it. The first two summands in (\ref{2ptactionz}) diverge as $\epsilon_{0}\rightarrow 0$ and they must to be subtracted altogether to get a finite result in the IR limit. Hence we will be left only with the contribution
\begin{equation}
{\cal S}=-\frac{1}{2\kappa^2}\int
d^d x \left. \left\{\left(\frac{\epsilon_0}{L}\right)^{1-d}\,\frac{\kappa^2}{4} H^j_i[z,\vec{x}]\partial_z H^i_j[z,\vec{x}]\right\}\right\vert_{z=\epsilon_0}.
\label{calS}
\end{equation}
This is precisely, modulo conventions, the result in (2.26) of the paper \cite{Arutyunov:1998ve}, where it is argued that (2.26) yields the correct two-point function of the energy-momentum tensor of the dual theory. Notice that our $H^j_i[z,\vec{k}]$, the Fourier transform of  $H^j_i[z,\vec{k}]$, is the same as $\bar{h}^i_j[z,\vec{k}]$ in  \cite{Arutyunov:1998ve}. Indeed, the latter  is traceless and transverse --see (2.21) of \cite{Arutyunov:1998ve}-- and its  actual value is given in (2.23) of \cite{Arutyunov:1998ve}, which is our (\ref{HIJresult}). Let us point out that to reach the conclusion just stated one may carry out the whole computation in momentum and see that the IR finite contribution to ${\cal S}$ in (\ref{calS})
reads
\begin{equation*}
{\cal S}_{\tiny{finite}}=C_{T}\idqd\idpd\,(2\pi)^d\delta(\vec{p}+\,\vec{q})\, h^{(b)}_{ij}(\vec{q})\Pi^{ij\,lm}\,(\vec{p})\,F(\vec{p})\,h^{(b)}_{lm}(\vec{p}),
\end{equation*}
where $C_{T}$ is a constant and
\begin{equation}
\begin{array}{l}
{\Pi^{ij\,lm}(\vec{p})=\frac{1}{2}\left(\pi^{il}(\vec{p})\pi^{jm}(\vec{p})+\pi^{im}(\vec{p})\pi^{jl}(\vec{p})\right)-\frac{1}{d-1}\pi^{ij}(\vec{p})\pi^{lm}(\vec{p}),}
\\[8pt]
{\pi^{ij}(\vec{p})=\delta^{ij}-\frac{p^i p^j}{p^2},}\\[8pt]
{F(p)= |\vec{p}|^d,\quad\text{if  $d$ is odd}\quad\text{and}\quad |\vec{p}|^d\ln|\vec{p}|,\quad\text{if  $d$ is even}.}
\label{skenderis}
\end{array}
\end{equation}
Taking two derivatives of ${\cal S}_{\tiny{finite}}$ with respect to $h^{(b)}_{ij}(\vec{p})$ yields, modulo a constant, the two-point correlation function of the energy-momentum tensor in momentum space found in \cite{Bzowski:2013sza} for general CFT. $F(\vec{p})$ in (\ref{skenderis}) can be read off from the on-shell action of a massless scalar field on Eclidean AdS --see \cite{Ammon:2015wua}.

Let us point out that our $H^j_i[z,\vec{k}]$ agrees with the bulk-boundary propagator used in \cite{Raju:2011mp, Albayrak:2019yve}. Indeed, the propagator in question is the solution in the axial gauge to the linearized Einstein equations for Dirichlet Boundary conditions and space-like momenta.

Let us now go back to the first two terms in (\ref{2ptactionz}) that we have subtracted to get an IR finite result. The corresponding contributions in General Relativity can be  obtained from equation (4.15) of \cite{Liu:1998bu} and they read
\begin{equation}
-\frac{1}{2\kappa^2}\int d^d x \,\frac{2(d-1)}{L}\left(\frac{\epsilon_0}{L}\right)^{-d}\left(1-\frac{\kappa^2}{4}h^i_j h^j_i\right).
\label{GRresult}
\end{equation}
Obviously, the integrand of (\ref{GRresult}) and the two first summands of  (\ref{2ptactionz})  are linear combinations of the same type of monomials, namely $1$ and $h^i_j h^j_i$, but with different coefficients. So these IR divergent contributions in General Relativity differ from those of our unimodular theory.

It has been shown in \cite{Liu:1998bu} that the IR divergences we have just quoted can be subtracted just by adding the term
\begin{equation*}
a\,\int d^d x \sqrt{g^{(b)}}
\end{equation*}
and choosing the coefficient $a$ appropriately.  One may wonder if the analogous term, namely
\begin{equation*}
\frac{c}{L}\,\int d^d x \sqrt{\hat{g}^{(b)}},
\end{equation*}
would do the job for unimodular gravity. The answer is no, for the expansion in (\ref{hgbdet}) yields the following contribution
\begin{equation*}
\frac{c}{L}\,\int d^d x\, \frac{\rho_0 d}{L}\,\Big[1-\frac{1}{4(d+1)}\kappa^2 h^i_j h^j_i\Big],
\end{equation*}
so that one can choose, eg, $c=2(1-d)$, to cancel the $h^i_j h^j_i$ summand  in (\ref{2ptaction}); but, then there remains an IR --ie, as $\rho_0\rightarrow\infty$-- divergent  contribution
\begin{equation*}
\int d^d x\, \frac{\rho_0 d}{L^2}\,(1-d),
\end{equation*}
which has to be subtracted anyway.

Summarizing, we have shown that, up to the quadratic order, the value of the on-shell classical action for our unimodular gravity differs from that of General Relativity by IR divergent contact terms --see (\ref{2ptactionz}) and (\ref{GRresult}). Hence, our unimodular theory differs from  General Relativity at the (IR) regularized level.  And yet, for the leading saddle point approximation to the  two-point contribution of the gravity field to $\ln Z_{gravity}[h^{(b)}_{ij}]$ in (\ref{saddlep}), a sensible subtraction of the IR divergences yields the same finite result for our unimodular theory as for General Relativity. So the equivalence between our unimodular gravity theory and General Relativity holds, in the case at hand, in a nontrivial way. Of course, the two-point correlation function of the energy-momentum tensor of the dual theory obtained from the on-shell classical gravity action in the leading approximation is the same for both unimodular theory and General Relativity.

\section{The three-point function}

Here we shall work out the contribution to $S_{\text{\tiny{UG}}}$ in (\ref{UGaction}) involving three $h_{\mu\nu}$, $h_{\mu\nu}$  being given in (\ref{finalresult}). We shall compare the contribution in question with that of General Relativity and draw conclusions.

The use of the algebraic package xAct \cite{xAct} and some very lengthy computations yields that the three-$h_{\mu\nu}$ contribution, say $S_{\text{\tiny{HEUG3}}}$, to
\begin{equation*}
 -\frac{1}{2\kappa^2}\int d^d x\int_{0}^{\rho_0}dw \,R[\hat{g}]
\end{equation*}
reads
\begin{equation}
S_{\text{\tiny{HEUG3}}}= S_{\text{\tiny{BulkUG3}}}\,+\, {\cal B}_{\text{\tiny{HEUG3}}},
\label{SHEUG3}
\end{equation}
where
\begin{equation}
\begin{array}{l}
{S_{\text{\tiny{BulkUG3}}}=
-\frac{\kappa}{2}\int
 d^d x \int_0^{\rho_0} dw\,\sqrt{\bar{g}}\Big\{\frac{d}{6 L^2}h^\mu_\lambda h^\lambda_\nu h^\nu_\mu+\frac{1}{4}h^{\mu\nu}\bar{\nabla}_\mu h_{\tau\sigma}\bar{\nabla}_\nu h^{\tau\sigma}
-\frac{1}{2}h^{\mu\tau}\bar{\nabla}_\tau h^{\nu\sigma}\bar{\nabla}_\sigma h_{\mu\nu}\Big\},}\\
{{\cal B}_{\text{\tiny{HEUG3}}}=-\frac{\kappa}{2}\int
 d^d x \int_0^{\rho_0} dw\,\bar{\nabla}_\lambda B^\lambda ,}\\[8pt]
{B^\lambda=\frac{d-3}{4(d+1)}h^{\mu\nu}h_{\nu\tau}\bar{\nabla}^\lambda h_\mu^\tau-\frac{1}{d+1}h^{\mu\nu}h^{\lambda\tau}\bar{\nabla}_\tau h_{\mu\nu}+h^{\mu\lambda}h^{\nu\tau}\bar{\nabla}_\tau h_{\mu\nu}+\frac{1}{2}h_{\mu\nu}h^{\nu\tau} \bar{\nabla}_\tau h^{\mu\lambda}.}
\label{B3}
\end{array}
\end{equation}
To obtain (\ref{SHEUG3}) and (\ref{B3}), the equations in (\ref{TTcovariant}) and (\ref{simpleeom}) are to be employed profusely.

Let us simplify the boundary contribution, ${\cal B}_{\text{\tiny{HEUG3}}}$,  to $S_{\text{\tiny{HEUG3}}}$ by imposing the axial gauge condition $h_{\mu0}[w,\vec{x}]=0$:
\begin{equation}
\begin{array}{l}
{{\cal B}_{\text{\tiny{HEUG3}}}\,=\,-\frac{\kappa}{2}\int
 d^d x \int_0^{\rho_0} dw\,\sqrt{\bar{g}}\, \bar{\nabla}_\lambda B^\lambda[w,\vec{x}]=-\frac{\kappa}{2}\int
 d^d x \left.\left[\sqrt{\bar{g}^{(b)}}\,\bar{n}_\lambda\, B^\lambda\right]\right\vert_{w=\rho_0}=}\\[8pt]
{\phantom{{\cal B}_{\text{\tiny{HEUG3}}}\,}-\frac{\kappa}{2}\int
 d^d x \left.\Big[\left(\frac{\rho_0 d}{L}\right)^2\, \frac{d-3}{4(d+1)}\,h^{ij}h_{i}^k \partial_0 h_{jk}-\frac{\rho_0 d}{L^2 }\frac{d-1}{d+1}h^i_jh^j_l h^l_i\Big]\right\vert_{w=\rho_0}=}\\[8pt]
 {\phantom{{\cal B}_{\text{\tiny{HEUG3}}}\,}-\frac{\kappa}{2}\int
 d^d x \left.\Big[\left(\frac{\rho_0 d}{L}\right)^2\, \frac{d-3}{4(d+1)}\,h^i_jh^j_l \partial_0 h^l_i-\frac{\rho_0 d}{L^2 }\Big(\frac{1}{2}\Big) h^i_jh^j_l h^l_i\Big]\right\vert_{w=\rho_0}=}
 \end{array}
 \label{BoundaryUG3}
\end{equation}
where $\bar{g}^{(b)}_{ij}$ and $\bar{n}^\lambda$ are given in (\ref{bbgmetric}) and (\ref{nbarvalue}), respectively. Recall that  $\partial_0\equiv\frac{\partial}{\partial w}$ and that $\bar{g}=1$.

To compute the three-$h_{\mu\nu}$ contribution coming from the unimodular Hawking-Gibbons-York action in (\ref{HGYaction}), the following results are needed
\begin{equation}
\begin{array}{l}
{\sqrt{\hat{g}^{(b)}[\rho_0,\vec{x}]}=\frac{\rho_0 d}{L}\left.\Big[1-\frac{1}{4(d+1)}\kappa^2 h^i_jh^j_i+\frac{1}{6(d+1)}\kappa^3 h^i_j h^j_l h^l_i\Big]\right\vert_{w=\rho_0}+o((h_{ij})^4),}\\[8pt]
{\hat{n}^{0}=\hat{n}^0[w,\vec{x}]=\frac{ w d }{L}\Big[1-\frac{1}{4(d+1)}\kappa^2 h^i_j h^j_i+\frac{1}{6(d+1)}\kappa^3 h^i_j h^j_l h^l_i\Big]+o((h_{ij})^4),}\\[8pt]
{ K[\rho_0,\vec{x}]=\partial_\mu \hat{n}^\mu[\rho_0,\vec{x}]=\partial_0\hat{n}^0[w,\vec{x}]=\frac{d}{L}\left.\Big[1-\frac{1}{4(d+1)}\kappa^2 h^i_j h^j_i+\frac{1}{6(d+1)}\kappa^3 h^i_j h^j_l h^l_i\Big]\right\vert_{w=\rho_0}+}\\[8pt]
{\phantom{K[\rho_0,\vec{x}]=}+\frac{\rho_0 d}{L}\left.\Big[-\frac{1}{2(d+1)}\kappa^2 h^i_j\partial_0 h^j_i+\frac{1}{2(d+1)}\kappa^3 h^i_j h^j_l \partial_0 h^l_i\Big]\right\vert_{w=\rho_0}+o((h_{ij})^4),}
\label{sundryresults}
\end{array}
\end{equation}
where $\hat{g}^{(b)}_{ij}$ has been defined in (\ref{hgbij}) and $\hat{n}^\mu=(\hat{n}^0,\vec{0})$ and $K[\rho_0,\vec{x}]$ have been introduced in the  paragraph beginning right below (\ref{hgbdet}). To obtain (\ref{sundryresults}) the conditions $h_{\mu 0}=0$ and $h=0$ must be imposed, recall that these
conditions are satisfied by our solution in (\ref{finalresult}).

Using the results in (\ref{sundryresults}), it can be shown that three-field contribution, $S_{\text{\tiny{HGY3}}}$, to the action in (\ref{HGYaction}) runs thus:
\begin{equation}
S_{\text{\tiny{HGY3}}}=-\frac{\kappa}{2}\int
 d^d x \left.\Big[\left(\frac{\rho_0 d}{L}\right)^2\, \frac{1}{d+1}\,h^i_jh^j_l \partial_0 h^l_i+\frac{\rho_0 d}{L^2 }\frac{2 d}{3(d+1)}h^i_jh^j_l h^l_i\Big]\right\vert_{w=\rho_0},
 \label{SHGY3}
\end{equation}

Let us introduce ${\cal B}_{\text{\tiny{UG3}}}$:
\begin{equation}
{\cal B}_{\text{\tiny{UG3}}}={\cal B}_{\text{\tiny{HEUG3}}}+S_{\text{\tiny{HGY3}}}=
-\frac{\kappa}{2}\int
 d^d x \left.\Big[\left(\frac{\rho_0 d}{L}\right)^2\,\Bigg( \frac{1}{4}\Bigg)\,h^i_jh^j_l \partial_0 h^l_i+\frac{\rho_0 d}{L^2 }\Bigg(\frac{ d-3}{6(d+1)}\Bigg) h^i_j h^j_l h^l_i\Big]\right\vert_{w=\rho_0},
\label{boundarycalUG3}
\end{equation}
where ${\cal B}_{\text{\tiny{HEUG3}}}$ and $S_{\text{\tiny{HGY3}}}$ are displayed in (\ref{BoundaryUG3}) and (\ref{SHGY3}), respectively.

We then conclude that the three-$h_{\mu\nu}$ contribution, $S_{\text{\tiny{UG3}}}$, to $S_{\text{\tiny{UG}}}$ in (\ref{UGaction}) is given by
\begin{equation}
S_{\text{\tiny{UG3}}}= S_{\text{\tiny{BulkUG3}}}\,+\,{\cal B}_{UG3},
\label{SUG3}
\end{equation}
where $S_{\text{\tiny{BulkUG3}}}$ and ${\cal B}_{UG3}$ can be found in (\ref{B3}) and (\ref{boundarycalUG3}), respectively.

Let us carry out a similar computation for General Relativity. To do so we shall need the following result obtained in the Appendix, namely, that, modulo a gauge transformation,  the solution, in the axial gauge $h_{\mu 0}=0$ and  having a well-defined limit as
$z\rightarrow \infty$, to the linearized General Relativity equations for Dirichlet boundary conditions  satisfies
\begin{equation}
h[z,\vec{x}]=\tilde{g}^{\mu\nu}h_{\mu\nu}[z,\vec{x}],\quad \tilde{\nabla}^\mu h_{\mu\nu}[z,\vec{x}]=0\quad\text{and}\quad \tilde{\Box}h_{\mu\nu}[z,\vec{x}]=-\frac{2}{L^2}h_{\mu\nu}[z,\vec{x}].
\label{TTGRcovariant}
\end{equation}
$\tilde{g}_{\mu\nu}$ is the Euclidean AdS metric with line element in (\ref{standardmetric}). The covariant derivative $\tilde{\nabla}_\mu$ is defined with regard to $\tilde{g}_{\mu\nu}$.

Let $S_{\text{\tiny{GR}}}$ be defined as follows
\begin{equation}
\begin{array}{l}
{ S_{\text{\tiny{GR}}}\,=\, S_{\text{\tiny{HEGR}}}\,+\, S_{\text{\tiny{HGYGR}}},}\\[8pt]
{S_{\text{\tiny{HEGR}}}=  -\frac{1}{2\kappa^2}\int d^d x\int^{\infty}_{\epsilon_0} dz \,\sqrt{g}\,\big(R[g_{\mu\nu}]+ \frac{d(d-1)}{L^2}\big),\quad
 S_{\text{\tiny{HGYGR}}}=-\frac{1}{2\kappa^2}\int d^d x\,\left.2\sqrt{g^{(b)}}K\right\vert_{z=\epsilon_0},}
 \label{GRaction}
 \end{array}
\end{equation}
where $g_{\mu\nu}=\tilde{g}_{\mu\nu}+\kappa h_{\mu\nu}$. The computation of the three-field contribution, $S_{\text{\tiny{HEGR3}}}$, to $S_{\text{\tiny{HEGR}}}$ yields
\begin{equation}
\begin{array}{l}
S_{\text{\tiny{HEGR3}}}= S_{\text{\tiny{BulkGR3}}}\,+\, {\cal B}_{\text{\tiny{HEGR3}}},
\label{SHEGR3}
\end{array}
\end{equation}
where
\begin{equation}
\begin{array}{l}
{S_{\text{\tiny{BulkGR3}}}= -\frac{\kappa}{2}\int
d^d x\int^{\infty}_{\epsilon_0} dz \sqrt{\tilde{g}}\Big\{\frac{d}{6L^2}h_\mu^\lambda h_{\nu\lambda} h^{\mu\nu}+\frac{1}{4}h^{\mu\nu}\tilde{\nabla}_\mu h_{\tau\s}\tilde{\nabla}_\nu h^{\tau\s}-\frac{1}{2}h^{\mu\tau}\tilde{\nabla}_\tau h^{\nu\s}\tilde{\nabla}_\s h_{\mu\nu}\Big\},}\\[8pt]
{{\cal B}_{\text{\tiny{HERG3}}}= -\frac{\kappa}{2}\int
d^d x\int^{\infty}_{\epsilon_0} dz\; \tilde{\nabla}_\lambda B_{\text{\tiny{GR3}}}^\lambda ,}\\[8pt]
{B_{\text{\tiny{GR3}}}^\lambda=-\frac{3}{4}h^{\mu\nu}h_{\nu\tau}\tilde{\nabla}^\lambda h_\mu^\tau-h^{\mu\nu}h^{\lambda\tau}\tilde{\nabla}_\tau h_{\mu\nu}+h^{\mu\lambda}h^{\nu\tau}\tilde{\nabla}_\tau h_{\mu\nu}+\frac{1}{2}h_{\mu\nu}h^{\nu\tau} \tilde{\nabla}_\tau h^{\mu\lambda}.}
\label{BGR3}
\end{array}
\end{equation}
To obtain (\ref{SHEGR3}) and (\ref{BGR3}) we have integrated by parts --keeping the boundary contributions-- and used (\ref{TTGRcovariant}).

The axial gauge condition $h_{\mu 0}=0$ and a little algebra leads to the conclusion that
\begin{equation}
{\cal B}_{\text{\tiny{HERG3}}}=-\frac{\kappa}{2}\int
 d^d x \left.\Big[\left(\frac{\epsilon_0}{L}\right)^{1-d}\, \left(\frac{3}{4}\right)\,h^i_jh^j_l \partial_z h^l_i+\left(\frac{\epsilon_0 }{L}\right)^{-d}\left(-\frac{1}{2 L}\right) h^i_jh^j_l h^l_i\Big]\right\vert_{z=\epsilon_0}.
 \label{SHERG3}
\end{equation}

It has been shown in \cite{Liu:1998bu} that
\begin{equation}
S_{\text{\tiny{HGYGR}}}= -\frac{1}{2\kappa^2}\int d^d x\,\left.\left(-2z\right)\frac{\partial}{\partial z}\sqrt{g^{(b)}[z,\vec{x}]}\right\vert_{z=\epsilon_0},
\label{lagata}
\end{equation}
where $S_{\text{\tiny{HGYGR}}}$ is defined in (\ref{GRaction}) and $g^{(b)}[z,\vec{x}]$ denotes the determinant of $g_{ij}[z,\vec{x}]$. Hence, by taking into account that
\begin{equation*}
{\sqrt{g^{(b)}[z,\vec{x}]}=\left(\frac{L}{z}\right)^d\Big[1-\frac{1}{4}\kappa^2 h^i_jh^j_i+\frac{1}{6}\kappa^3 h^i_j h^j_l h^l_i\Big]+o((h_{ij})^4),}
\end{equation*}
we obtain that the three-$h_{ij}$ contribution to $S_{\text{\tiny{HGYGR}}}$ in (\ref{lagata}) reads
\begin{equation}
S_{\text{\tiny{HGYGR3}}}=-\frac{\kappa}{2}\int
 d^d x \left.\Big[\left(\frac{\epsilon_0}{L}\right)^{1-d}\, \left(-1\right)\,h^i_jh^j_l \partial_z h^l_i+\left(\frac{\epsilon_0 }{L}\right)^{-d}\left(\frac{d}{3 L}\right) h^i_jh^j_l h^l_i\Big]\right\vert_{z=\epsilon_0}.
\label{SHGYGR3}
\end{equation}
Putting it all together we conclude that the three-field contribution, $ S_{\text{\tiny{GR3}}}$, to  $S_{\text{\tiny{GR}}}$ in (\ref{GRaction}) is given by
\begin{equation}
S_{\text{\tiny{GR3}}}=S_{\text{\tiny{BulkGR3}}}\,+\,{\cal B}_{\text{\tiny{GR3}}},
\label{SGR3}
\end{equation}
where $S_{\text{\tiny{BulkGR3}}}$ is displayed in (\ref{BGR3}) and
\begin{equation}
\begin{array}{l}
{{\cal B}_{\text{\tiny{GR3}}}={\cal B}_{\text{\tiny{HEGR3}}}\,+\,S_{\text{\tiny{HGYGR3}}}=}\\[8pt]
{-\frac{\kappa}{2}\int
 d^d x \left.\Big[\left(\frac{\epsilon_0}{L}\right)^{1-d}\, \left(-\frac{1}{4}\right)\,h^i_jh^j_l \partial_z h^l_i+\left(\frac{\epsilon_0 }{L}\right)^{- d}\left(\frac{2d-3 }{6 L}\right) h^i_jh^j_l h^l_i\Big]\right\vert_{z=\epsilon_0}.}
\label{BoundGR3}
\end{array}
\end{equation}
The values of ${\cal B}_{\text{\tiny{HEGR3}}}$ and $S_{\text{\tiny{HGYGR3}}}$ can be found in (\ref{SHERG3}) and (\ref{SHGYGR3}), respectively.

We may compare now the three-field contribution $S_{\text{\tiny{UG3}}}$ in (\ref{SUG3}) with the three-field contribution $S_{\text{\tiny{GR3}}}$ in (\ref{SGR3}). But before we make that comparison, let us point out a fact regarding the $h^i_j[z,\vec{x}]$ field which ${\it i})$ solves the linearized General Relativity equations for the metric in (\ref{standardmetric}), ${\it ii})$  satisfies Dirichlet boundary conditions and ${\it iii})$ has a well-defined limit as $z\rightarrow\infty$. The fact is that $h^i_j[z,\vec{x}]=H^i_j[z,\vec{x}]$, where the Fourier transform of $H^i_j[z,\vec{x}]$ is given in (\ref{HIJresult}) and (\ref{finalresult}). The reader should consult the Appendix for details.

It is plane that the change of variables $z\rightarrow w$ defined in (\ref{wcoord}) turns $S_{\text{\tiny{BulkGR3}}}$, in (\ref{BGR3}), into
$S_{\text{\tiny{BulkUG3}}}$  in (\ref{B3}). However, if we apply the change of variables we have just mentioned to ${\cal B}_{\text{\tiny{GR3}}}$ in (\ref{BoundGR3}), we get
\begin{equation}
{\cal B}_{\text{\tiny{GR3}}}=
-\frac{\kappa}{2}\int
 d^d x \left.\Big[\left(\frac{\rho_0 d}{L}\right)^{2}\, \left(\frac{1}{4}\right)\,h^i_jh^j_l \partial_0 h^l_i+
 \left(\frac{\rho_0 d }{L^2}\right)
 \left(\frac{2d-3 }{6 }\right) h^i_jh^j_l h^l_i\Big]\right\vert_{w=\rho_0},
\label{BGR3w}
\end{equation}
where $\rho_0= \frac{L^{d+1}}{d}(\epsilon_0)^{-d}$, $\partial_0=\frac{\partial}{\partial w}$ and $h^i_j=h^i_j[w,\vec{x}]$; the Fourier transform of $h^i_j[w,\vec{x}]$ being given in (\ref{finalresult}).

Obviously, ${\cal B}_{\text{\tiny{GR3}}}$ in (\ref{BGR3w}) and ${\cal B}_{\text{\tiny{UG3}}}$ in (\ref{boundarycalUG3}) are not equal, the difference
coming from the IR divergent contact term
\begin{equation}
\int d^d x \left. \Big[\left(\frac{\rho_0 d }{L^2}\right)
  h^i_jh^j_l h^l_i\Big]\right\vert_{w=\rho_0}= \int d^d x\left(\frac{\rho_0 d }{L^2}\right)
  h^{(\tiny{TT})\,i}_j[\vec{x}]h^{(\tiny{TT})\,j}_l[\vec{x}] h^{(\tiny{TT})\,l}_i[\vec{x}],
\label{contact3}
\end{equation}
where $h^{(\tiny{TT})\,i}_j[\vec{x}]$ has  $h^{TT\,i}_j[\vec{k}]$ in (\ref{hTTdecomp}) as Fourier transform.
This term --since it is a contact term-- does not contribute to value of the three-point correlation function of the energy-momentum tensor of the dual field theory. We see again the same picture as  for the two-point contribution discussed in the previous section. Indeed, the three-field contribution to righthand side of (\ref{saddlep}) in unimodular gravity is not the same as in General Relativity when the IR regulator is in place. However, the difference is  an IR divergent contact term which  does not contribute to the value of the three-point correlation functions of the energy-momentum tensor of the dual field theory. Of course, the subtraction of the term in question to get an IR finite value for the right hand side of (\ref{saddlep}) will make unimodular gravity fully equivalent to General Relativity as far as our results are concerned. This  equivalence arising in a nontrivial way, though.

\section{Summary and Conclusions}

The formulation of theory of unimodular gravity put forward in \cite{Alvarez:2006uu, Alvarez:2005iy, Alvarez:2015sba} has the nice feature that transverse diffeomorphims and Weyl transformations are the gauge symmetries of the theory. We have started the study of the properties of this formulation of unimodular gravity from the gauge/gravity duality point of view. We do so by computing --at the lowest order-- the IR regularized two- and three-point $h_{\mu\nu}$ contributions to the on-shell classical gravity action for an Euclidean AdS background. We have shown that these two- and three- point contributions do not agree with the corresponding contributions in General Relativity due to IR divergent contact terms --see (\ref{2ptactionz}) and (\ref{GRresult}), on the one hand, and (\ref{boundarycalUG3}), (\ref{BGR3w}) and (\ref{contact3}), on the other. However, once those IR divergent terms are subtracted our unimodular theory and General Relativity yield the same IR finite result. The subtraction in question does not modify the value of the corresponding correlation functions of the energy-momentum tensor of the dual field theory. So, we conclude that, as far as our computations can tell, our unimodular gravity theory and General Relativity are equivalent in the sense that they have the same dual boundary field theory. Of course, we have shown that this equivalence emerges in a nontrivial way. Whether the equivalence in question will still hold for higher-point functions and/or when one-loop corrections are taken into account is an open problem. A problem which is worth studying.

\section{Acknowledgments} We thank E. \'Alvarez for continuous and illuminating discussions.
The work by C.P. Martin has been financially supported in part by the Spanish MICINN through grant {\color{green} PGC2018-095382-B-I00}.

\section{Appendix}

In this Appendix we shall discuss how to find a suitable solution to the linearized General Relativity equations
\begin{equation}
\begin{array}{l}
{\frac{1}{2}\tilde{\Box}h_{\mu\nu}-\frac{1}{2}\tilde{g}_{\mu\nu}\tilde{\Box}h-\frac{1}{2}\tilde{\nabla}_\mu\tilde{\nabla}_\lambda h^\lambda_\nu-\frac{1}{2}\tilde{\nabla}_\nu\tilde{\nabla}_\lambda h^\lambda_\mu+\frac{1}{2}\tilde{g}_{\mu\nu}\tilde{\nabla}_\tau\tilde{\nabla}_\sigma h^{\tau\sigma}+\frac{1}{2}\tilde{\nabla}_\mu\tilde{\nabla}_\nu h-}\\[8pt]
{+\frac{1}{L^2} h_{\mu\nu}+\frac{(d-2)}{2L^2}\tilde{g}_{\mu\nu} h=0,}
\label{linGR}
\end{array}
\end{equation}
where $\tilde{g}_{\mu\nu}$ is the metric with line element in (\ref{standardmetric}) and all covariant derivatives are defined with regard to $\tilde{g}_{\mu\nu}$. Let us recall that given an arbitrary real vector field, $U^\mu[z,\vec{x}]$, the previous equation is invariant the gauge transformations
\begin{equation*}
\delta h_{\mu\nu}=\tilde{\nabla}_\mu U_\nu+\tilde{\nabla}_\nu U_\mu.
\end{equation*}
 We shall obtain the solution to (\ref{linGR}) in the axial gauge, $h_{\mu0}=0$, which satisfies appropriate Dirichlet boundary conditions and has a well-defined limit as $z\rightarrow \infty$. The domain where (\ref{linGR}) will be solved is $\{(z,\vec{x}); \epsilon_{0}<z<\infty,\vec{x}\in\rm{I\!R}^d\}$, with boundary at $z=\epsilon_0$. What we shall find is that the solution in question, say $h_{ij}[z,\vec{x}]$, is such that its Fourier transform $h_{ij}[z,\vec{k}]$ is, modulo a gauge transformation, equal to $\tilde{g}_{il}H^l_j[z,\vec{k}]$, $H^l_j[z,\vec{k}]$
being given in (\ref{HIJresult}); the  gauge transformation preserving the axial gauge condition. This means that this is the solution --see (\ref{finalresult})-- we found for the linearized unimodular gravity equation in (\ref{linearisedugeq}) expressed in terms of the coordinate $z$ instead of the coordinate $w$ in (\ref{wcoord}). Notice, though, that this result is nontrivial, for (\ref{linGR}) and (\ref{linearisedugeq}) are quite different. It is important to stress that the solution to (\ref{linGR}) that we shall find
satisfies
\begin{equation*}
h[z,\vec{x}]=\tilde{g}^{\mu\nu}h_{\mu\nu}[z,\vec{x}],\quad \tilde{\nabla}^\mu h_{\mu\nu}[z,\vec{x}]=0\quad\text{and}\quad \tilde{\Box}h_{\mu\nu}[z,\vec{x}]=-\frac{2}{L^2}h_{\mu\nu}[z,\vec{x}],
\end{equation*}
for this was used in our computations of the General Relativity three-field contributions to the right hand side of (\ref{saddlep}).

Let us point out that our result is not new. In \cite{Raju:2011mp} --see its eq. (2.43)-- it is stated that the axial gauge bulk-boundary propagator for the gravitational field for space-like momenta is given by $\tilde{g}_{il}H^l_j[z,\vec{k}]$, where  $H^l_j[z,\vec{k}]$ is displayed in (\ref{HIJresult}).
This bulk-boundary propagator is no other thing that the solution to the linerized General Relativity equations with Lorenztian signature  for space-like momenta and for the boundary conditions and behaviour in the AdS interior stated in the previous paragraph. Of course, this solution yields the solution of the corresponding equations with Euclidean signature --ie, the equations in (\ref{linGR}). Indeed, one just has to replace in the former solution the space-like $k^2$  with $k^2$ defined with Euclidean signature; bear in mind that we are using the most plus Lorentz metric.

Although, as we have discussed in the previous paragraph, the final result presented in this Appendix is not new, we think that the analysis we shall display  below will be helpful.

The Fourier transform  with regard to $\vec{x}$ of the $00$, $0j$ and $ij$ components of the equation in (\ref{linGR}) read
\begin{equation}
(2(d-1)\breve{h}+z^2k^2) \breve{h}+(d-1)z\breve{h}'-z^2k^ik^jh_{ij}=0
\label{338}
\end{equation}
\begin{equation}
2(k^l h_{lj}-k_j \breve{h})+z(k^l h_{lj}'-k_j \breve{h}')=0
\label{339}
\end{equation}
and
\begin{equation}
\begin{array}{l}
{-z^2 h_{ij}''+(d-5)z h_{ij}'+ (2(d-2)+k^2 z^2)h_{ij}-z^2\big[k^lk_i h_{lj}+k^lk_j h_{li}\big]+z^2\delta_{ij}k^lk^m h_{lm}+}\\[8pt]
{\delta_{ij} z^2\breve{h}''+(5-d)\delta_{ij} z\breve{h}'+(-2 (-2 + d) - k^2 z^2)\delta_{ij} + k_i k_j z^2)\breve{h}=0,}
\label{340}
\end{array}
\end{equation}
respectively. $h_{ij}$ is a function of $z$ and the Fourier momentum $\vec{k}$. $\breve{h}\equiv \delta^{ij} h_{ij}$.

The general solution to (\ref{339}) reads
\begin{equation}
k^l h_{lj}-k_j \breve{h}= \frac{v_j[\vec{k}]}{z^2},
\label{341}
\end{equation}
where $v_j[\vec{k}], j=1..d$ are integration constants. Substituting (\ref{341}) in (\ref{338}), one gets
\begin{equation*}
2(d-1)\breve{h}+(d-1)z\breve{h}'- \vec{k}\cdot \vec{v}[\vec{k}]=0,
\end{equation*}
whose general solution is the following
\begin{equation}
\breve{h}[z,\vec{k}]=\frac{C[\vec{k}]}{z^2}+\frac{1}{2(d-1)}\vec{k}\cdot \vec{v}[\vec{k}],
\label{342}
\end{equation}
where $C[\vec{k}]$ is another integration constant.

Now, since $z=\infty$ corresponds only to a point of Euclidean AdS and we want $h_{ij}[z,\vec{x}$] to have a well-defined --ie, independent of $\vec{x}$ -- limit as $z\rightarrow\infty$, we must demand that
\begin{equation}
\vec{k}\cdot \vec{v}[\vec{k}]=0.
\label{keyeq}
\end{equation}
Indeed, from (\ref{342}), one gets $\lim_{z\rightarrow\infty}\breve{h}[z,\vec{x}]=\frac{1}{2(d-1)}\vec{\partial}\cdot \vec{v}[\vec{x}]$, where $\vec{v}[\vec{x}]$ has $\vec{v}[\vec{k}]$ as Fourier transform. Hence, we must demand that $\vec{\partial}\cdot \vec{v}[\vec{x}]=A$, $A$ being a constant, if we want the large $z$ limit of $h_{ij}[z,\vec{x}]$ to be independent of $\vec{x}$. But, $A$ must be equal to zero, for $\vec{v}[\vec{x}]$ should vanish fast enough as $|\vec{x}|\rightarrow\infty$  --we are assuming that $\vec{v}[\vec{x}]$ has Fourier transform. $\vec{\partial}\cdot \vec{v}[\vec{x}]=0$ implies that its Fourier transform, $\vec{k}\cdot \vec{v}[\vec{k}]$, vanishes.

Let us take stock. What we have obtained so far is that
\begin{equation}
\breve{h}[z,\vec{k}]=\frac{C[\vec{k}]}{z^2},\quad\quad k^l h_{lj}=k_j \breve{h}+ \frac{v_j[\vec{k}]}{z^2}=\frac{1}{z^2}(v_j[\vec{k}]+k_j C[\vec{k}]).
\label{343}
\end{equation}
Recall that $\breve{h}\equiv \delta^{ij}h_{ij}$.

Let us introduce $h^{\tiny{part}}_{ij}[z,\vec{k}]$:
\begin{equation}
h^{\tiny{part}}_{ij}[z,\vec{k}]=\frac{1}{k^2 z^2}(k_i {\cal V}_j[\vec{k}]+k_j {\cal V}_i[\vec{k}]), \quad {\cal V}_j[\vec{k}]=v_j[\vec{k}]+\frac{1}{2} k_j C[\vec{k}].
\label{344}
\end{equation}
Notice that
\begin{equation}
\breve{h}^{\tiny{part}}[z,\vec{k}]=\frac{C[\vec{k}]}{z^2},\quad\quad k^l h^{\tiny{part}}_{lj}=\frac{1}{z^2}(v_j[\vec{k}]+k_j C[\vec{k}]),
\label{345}
\end{equation}
for (\ref{keyeq}) holds. But there is more: $h^{\tiny{part}}_{ij}[z,\vec{k}]$ solves (\ref{338}), (\ref{339}) and (\ref{340}), as can be seen by just substituting (\ref{344}) in those equations. This result is not surprising though, for $h^{\tiny{part}}_{ij}[z,\vec{x}]$ can be recast as gauge transformation that preserves the axial gauge condition $h_{0\mu}[z,\vec{x}]=0$. Indeed, let us define $\Theta_\mu[z,\vec{x}]$ as follows
\begin{equation}
\begin{array}{l}
{\Theta_\mu[z,\vec{x}]=(\Theta_0[z,\vec{x}],\Theta_i[z,\vec{x}]),}\\[8pt]
{\Theta_0[z,\vec{x}]=0,\quad  \Theta_i[z,\vec{x}]=\frac{i}{z^2}\int\frac{d^d k}{(2\pi)^d}\;e^{-i k\cdot x} \frac{{\cal V}_i [\vec{k}]}{k^2}.}
\label{gaugefield}
\end{array}
\end{equation}
Then the following gauge transformation
\begin{equation}
\tilde{\nabla}_\mu\Theta_{\nu}+\tilde{\nabla}_\nu \Theta_{\mu},
\label{gaugettrans}
\end{equation}
where the covariant derivative is defined with regard to the metric $\tilde{g}_{\mu\nu}$ with line element in (\ref{standardmetric}), is such that
\begin{equation*}
\begin{array}{l}
{\tilde{\nabla}_0\Theta_0=0,\quad \tilde{\nabla}_0\Theta_i+\tilde{\nabla}_i \Theta_0=0,}\\[8pt]
{\tilde{\nabla}_i\Theta_j+\tilde{\nabla}_j \Theta_i=\int\frac{d^d k}{(2\pi)^d}\;e^{-i k\cdot x}\,\frac{1}{z^2 k^2}(k_i {\cal V}_j[\vec{k}]+k_j {\cal V}_i[\vec{k}]).}
\end{array}
\end{equation*}
This last equation  is (\ref{344}).

Next, let express $h_{ij}[z,\vec{k}]$, a solution to (\ref{338}), (\ref{339}) and (\ref{340}) satisfying (\ref{343}) for given ${\cal V}_i[\vec{k}]$ and $C[\vec{k}]$, as follows:
\begin{equation}
h_{ij}[z,\vec{k}]\,=\,h_{ij}^{tt}[z,\vec{k}]+h_{ij}^{\tiny{part}}[z,\vec{k}].
\label{httdef}
\end{equation}
$h_{ij}^{\tiny{part}}[z,\vec{k}]$ is defined in (\ref{344}). It follows from (\ref{343}) and (\ref{345}) that
\begin{equation}
\breve{h}^{tt}[z,\vec{k}]=0,\quad\quad k^l h^{tt}_{lj}=0,
\label{transversGR}
\end{equation}
where $\breve{h}^{tt}[z,\vec{k}]\equiv \delta^{ij}h_{ij}^{\tiny{part}}[z,\vec{k}]$.

Substituting (\ref{httdef}) in (\ref{338}) and (\ref{339}), one sees that they are trivially satisfied. But the substitution of (\ref{httdef}) in (\ref{340}) yields the following equation
\begin{equation}
z^2 h_{ij}^{tt\,''}[z,\vec{k}]-(-5+d) z h_{ij}^{tt\,'}[z,\vec{k}]-(-4 + 2 d + k^2 z^2) h_{ij}^{tt}[z,\vec{k}]\,=\,0.
\label{odfbyhtt}
\end{equation}
We have met this equation already: it is equation (\ref{simpleeomij}). Hence, we know --see analysis below (\ref{simpleeomij})-- that the general solution to (\ref{odfbyhtt}) which has a well-defined limit as $z\rightarrow\infty$ and satisfy Dirichlet boundary condition at $z=\epsilon_0$ reads
\begin{equation*}
h_{ij}^{tt}[z,\vec{k}]=\left(\frac{L}{z}\right)^2\,h^{tt\,i}_{j}[z,\vec{k}],\quad h^{tt\,i}_{j}[z,\vec{k}]=H^i_j[z,\vec{k}],
\end{equation*}
where $H^i_j[z,\vec{k}]$ is given in (\ref{HIJresult}). Let us stress that the previous equation has been of paramount importance to our discussion in sections 4 and 5.

Now, it is not difficult to see that (\ref{transversGR}) can be recast as follows
\begin{equation*}
\tilde{g}^{\mu\nu}h_{\mu\nu}^{tt}=0,\quad \tilde{\nabla}^{\mu}h_{\mu\nu}^{tt}=0,
\end{equation*}
where $h_{0\mu}$ is by definition equal to zero. Substituting the previous to equation in (\ref{linGR}), one gets
\begin{equation*}
\tilde{\Box}h^{tt}_{\mu\nu}=-\frac{2}{L^2}.
\end{equation*}

Let us finally point out that $h_{ij}[z,\vec{x}]$, as obtained from its Fourier transform in (\ref{httdef}), differs from $h_{ij}^{tt}[z,\vec{x}]$ by the gauge transformation in (\ref{gaugettrans}) and (\ref{gaugefield}); this gauge transformation preserves the axial gauge condition.

\end{document}